\documentclass[preprint]{revtex4}
\usepackage[latin1]{inputenc}
\usepackage{graphicx,epstopdf}
\usepackage{subfigure}
\usepackage{epsfig}
\usepackage{bigints}
\usepackage[usenames]{color}

\begin{document}

\title{Bubble dynamics and the quark-hadron phase transition in nuclear collisions}

\author{D. A. Foga\c{c}a\dag\,, S. M. Sanches Jr.\dag\,, R. Fariello\dag\, and \, F. S. Navarra\dag}
\address{\dag\ Instituto de F\'{\i}sica, Universidade de S\~{a}o Paulo\\
Rua do Mat\~ao Travessa R, 187, 05508-090, S\~{a}o Paulo, SP, Brazil}

\begin{abstract}
We study the nucleation of a quark gluon plasma (QGP) phase in a hadron gas at  low temperatures and high baryon densities. This  kind of process will
presumably happen very often in nuclear collisions at FAIR and NICA.  When the appropriate energy densities (or baryon densities) and temperatures are reached
the conversion of one phase into
another is not instantaneous. It is a complex process, which involves the nucleation of bubbles of the new phase. One important element of this transition process
is the rate of growth of a QGP bubble. In order to estimate it  we solve the Relativistic Rayleigh$-$Plesset equation which governs the dynamics
of a relativistic spherical  bubble in a  strongly interacting medium. The baryon rich hadron gas is represented by the nonlinear Walecka model  and the QGP
is described by the MIT bag model and also by a mean field model of QCD.

\end{abstract}

\maketitle

\section{Introduction}

It is now believed that a quark gluon plasma (QGP) is formed at high temperatures in the high energy heavy ion collisions performed at the RHIC and LHC colliders.
Presumably the QGP will also be formed at low temperatures in  collisions performed in the future at FAIR \cite{fair} and NICA \cite{nica}.
After its formation the QGP expands, cools and hadronizes, creating the observed hadrons.  It is a complex process and some of its aspects need to
be better understood. One of these aspects is  the transition from  compressed hadronic matter, which we will call hadron gas (HG) to  quark-gluon plasma
at the initial stage of the  collisions at lower energies and high baryon densities.

In most of the existing hydrodynamical codes  this transition is performed just
by following the  behavior of some variables, such as, for example, the energy density, $\varepsilon$. During a collision, the regions where the energy
density exceeds the critical value, $\varepsilon_c$,  are assumed to be made of QGP and the corresponding equation of state starts to be used. In particular, the
transition is assumed to occur instantaneously, as soon as some criterion is fulfilled  (such as $\varepsilon > \varepsilon_c$). However we might expect that even when
this condition is satisfied the system needs some time to convert itself to the other phase. As we will discuss below, nucleation theory can be used to
incorporate this transition time in the theoretical description of these nuclear collisions.

The matter produced at RHIC and LHC is at high temperature and small chemical potential (and baryon density). A few years ago, lattice QCD studies established
that, under these conditions the  quark-hadron phase transition is actually a crossover. On the other hand, in the forthcoming nuclear collisions to be performed at
FAIR \cite{fair} and NICA \cite{nica}  the temperature is expected to be low and the chemical potential is expected to be high.
Indeed,  in these accelerators, in contrast
to what happens at the RHIC and at the LHC, the stopping power is large and most of the projectile and target baryons get trapped in the dense region. The  QGP, when
formed, will be surrounded by compressed nuclear matter. Model calculations strongly suggest that, at low temperatures and high baryon densities,
the quark-hadron conversion is a first order phase transition, for which the nucleation theory should apply.

Nucleation (or bubble formation) of a quark phase in  hadronic matter has been extensively studied in the context of compact stars and also in relativistic heavy
ion  physics,  where we may have both the nucleation of a QGP bubble \cite{lugones,heisel,bomba,bombando,madsen} in a hadron gas and also  the nucleation of  a
HG bubble in the QGP \cite{kapusta,madsen2}.
Most of the recent works on the subject were devoted to improve  the calculation of the surface tension, free energy, critical radius and nucleation rate \cite{pinto}.
In the context of heavy ion collisions there is an extensive literature on  the nucleation process \cite{bub1,mishu}.
However, apart from Refs. \cite{bub2,bub3}, there are not many other
studies about the space-time evolution of the bubbles.   This evolution is described by differential equations, all of them based on variants of the Rayleigh
equation \cite{ray} and, in particular,  the Rayleigh-Plesset (RP) equation \cite{ray2}. Nucleation theory and its differential equations can be applied to the
nuclear collisions to be performed at FAIR   to estimate the transition time from the initial HG to the QGP. This transition time depends on several variables,
such as the probability of bubble formation, the number of produced bubbles, their sizes and the velocity with which the bubbles expand (or collapse).  This last
aspect is the main subject of this work.

The collision dynamics at FAIR is very complex and is the subject of the whole part III of the  CBM physics
book \cite{fair}.  The models described  there are very detailed  and also very different from each other. In this context, we believe that it is interesting
to have a simple theoretical tool to help in estimating  the transition time discussed above. To this end, we can use the relativistic Rayleigh-Plesset equation,
as it will be explained below.

\begin{figure}
\begin{center}
\includegraphics[width=0.43\textwidth]{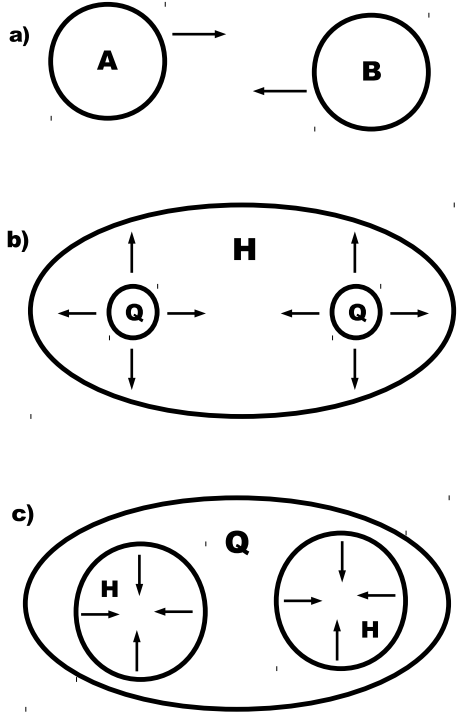}
\end{center}
\caption{Schematic picture of the initial stage  of a nuclear collision at intermediate energies. a)
Two nuclei collide. b) Bubbles of QGP (Q) start to form and grow in a hadron medium (H).
c) The remaining hadronic matter is in bubbles which shrink. In the end the QGP is formed.}
\label{fig1}
\end{figure}

A qualitative sketch of a nucleus-nucleus collision at intermediate energies  is depicted in Fig. \ref{fig1}.  Two nuclei collide (Fig. 1a). After  the collision
a compound system of hadronic matter (H) is formed which, under strong compression and at low temperatures,  goes from the hadronic (H) phase to the QGP (Q) phase.
The system is at high density ($\rho_B > \rho_c$) and at high pressure, the hadronic phase is unstable and  bubbles of  QGP may form and grow. In Fig. 1b) we see
QGP bubbles growing and  filling the whole system.  Later, the plasma phase is dominant, except  for  some regions where the hadronic phase persists. These regions are
represented by  bubbles of  the hadronic phase, which start to shrink.  This is shown in Fig. 1c). Actually the study of the growth of the QGP bubbles would be enough
to give an estimate of the transition time. For completeness we will also study the evolution and collapse of the hadron gas bubbles.

In the literature related to the  Rayleigh-Plesset equation, typically one speaks of a bubble of gas (G) immersed in a liquid (L). In the context of the present work
there is already a gas (the hadron gas) and hence, to avoid confusion we will use a different notation. All the quantities which refer to the matter inside a bubble
will be labeled with ``$in$'', such as the pressure, $p_{in}$,  and energy density, $\varepsilon_{in}$. The quantities which refer to the  matter outside
(external to) the bubble,  will be labeled with ``$ex$'', such as   $p_{ex}$ and  $\varepsilon_{ex}$.

Following the spirit of Refs. \cite{bub2} and  \cite{bub3}  but considering a  different physical scenario, in this work  we study the evolution of both HG and
QGP bubbles in warm and compressed hadronic matter. We find the solutions of the  relativistic Rayleigh-Plesset (RRP) equation derived in \cite{rrp} for various
plausible choices of parameters. These solutions will help us in determining the time scales associated with the hadron-to-quark  phase transition and QGP formation.

The work is organized as follows. In Section II we  generalize the RRP equation and the definitions of the involved quantities to the finite
temperature case. The detailed derivation is shown in the Appendix. We also present the
non-relativistic version of the Rayleigh-Plesset equation and its analytical solution. In Section III we show the equations of state which we are using.
In Section IV we show the numerical solutions of the RRP equation and discuss their implications and in Section V  we present some conclusions.

\section{Relativistic bubble evolution equation}

\subsection{Rayleigh-Plesset equation in relativistic fluid dynamics}

The Rayleigh-Plesset equation can be derived in different ways. It describes  the  motion of a spherical bubble filled with a medium (in) immersed in another medium
(ex). Here we follow the notation, the conventions ($\hbar=c=k_{B}=1$) and the approximations made in Ref. \cite{rrp}, where the  relativistic
Rayleigh$-$Plesset equation  was, to the best of our knowledge, derived for the  first time. In the Appendix we extend the formalism to  finite temperatures, in which the RRP equation  reads:
\begin{equation}
{\frac{d}{dt}} \, \Big[ (I_{1}+I_{2}) R^{3}\dot{R}  \Big] \,  +  \, (I_{1}-2I_{2}) \, R^{2}\, {\dot{R}}^{\,2} \, =  \, F
\label{rrp}
\end{equation}
where the integrals are given by (with $x=r/R$):
\begin{equation}
I_{1} \equiv \int_{0}^{1} dx \, x^{4} \,\bigg(\varepsilon_{in} +  {p_{in}}^{eff}  \bigg){\gamma_{in}}^{2}
\label{prei1}
\end{equation}
\begin{equation}
I_{2} \equiv \int_{1}^{\infty}{\frac{dx}{x^{2}}} \,\bigg(\varepsilon_{ex} +  {p_{ex}}^{eff}  \bigg){\gamma_{ex}}^{2}
\label{prei2}
\end{equation}
where the $\varepsilon$ and $p$ are the energy density and the pressure respectively. The  ``effective pressures'' are given by:
\begin{equation}
{p_{ex}}^{eff} \equiv p_{ex} -T_{ex}\,{\frac{\partial p_{ex}}{\partial T}}{\Big{|}}_{T=T_{ex}}
\hspace{1.5cm} \textrm{and} \hspace{1.5cm}
{p_{in}}^{eff} \equiv p_{in} -T_{in}\,{\frac{\partial p_{in}}{\partial T}}{\Big{|}}_{T=T_{in}}
\label{effectives}
\end{equation}
with $T_{ex}$ being the temperature of the external medium  and $T_{in}$ the temperature inside the bubble.

In the derivation of  (\ref{rrp}) the following Hubble-like velocity profile was assumed:
\begin{equation}
v_{in}(r)={\frac{r}{R}}\dot{R}
\hspace{0.8cm} \textrm{for} \hspace{0.8cm} 0<r<R
\label{fieldvg}
\end{equation}
and
\begin{equation}
v_{ex}(r)={\frac{R^{2}(R_{\infty}^{3}-r^{3})}{r^{2}(R_{\infty}^{3}-R^{3})}}\dot{R}
\hspace{0.8cm} \textrm{for} \hspace{0.8cm} R<r<R_{\infty}
\label{fieldvl}
\end{equation}
where $\dot{R}$ denotes the time derivative of the bubble radius, $R(t)$, and $R_{\infty}$  is the radius of the fluid, assumed to be very large.
These fields are used in the Lorentz factors:
\begin{equation}
\gamma_{in}={\frac{1}{\sqrt{ 1-\Bigg({\frac{r}{R}}\dot{R}\Bigg)^{2}}}}
\hspace{1.6cm} \textrm{and} \hspace{1.6cm}
\gamma_{ex}={\frac{1}{\sqrt{1
-\Bigg({\frac{R^{2}(R_{\infty}^{3}-r^{3})}{r^{2}(R_{\infty}^{3}-R^{3})}}
\dot{R}\Bigg)^{2}}}}
\label{lorentzfactors}
\end{equation}
The function $F$ is given by:
\begin{equation}
F \equiv -R^{2} \, \Bigg[ (\varepsilon_{in}-\varepsilon_{ex}){\Big{|}}_{R}+\bigg(\varepsilon_{ex} +p_{ex}-T\,{\frac{\partial p_{ex}}{\partial T}}\bigg){\Big{|}}_{\infty} \,
\Bigg]
+{\frac{3}{R}}\int_{0}^{R} dr \, r^{2} \,\bigg(\varepsilon_{in} +p_{in}-T\,{\frac{\partial p_{in}}{\partial T}}\bigg)
\label{efe}
\end{equation}
In what follows we will focus on the case where the phase boundary (i.e., the bubble surface) moves slowly compared to the
sound velocity of the medium $(\dot{R} << c_s \simeq 0.4 - 0.6)$ \cite{bombando}. Thus  the system can adjust adiabatically to the changes of the
bubble radius and each phase retains its pressure, which can be considered constant, i.e., independent of the  bubble radius.
In this approximation, in each phase $\varepsilon$ and $p$ are independent of the radial coordinate.
The energy density for each phase has then the following form:
\begin{equation}
\varepsilon_{ex}(r={R})=\varepsilon_{ex}(r\rightarrow{\infty})=\varepsilon_{ex} \hspace{0.9cm} \textrm{and} \hspace{0.9cm}
\varepsilon_{in}(r={R})=\varepsilon_{in}(r\rightarrow 0)=\varepsilon_{in}
\label{phasesrels}
\end{equation}
Inserting (\ref{phasesrels}) into (\ref{efe}) we obtain:
\begin{equation}
F = \Bigg[ (p_{in}-p_{ex})-\bigg(T_{in}\,{\frac{\partial p_{in}}{\partial T}}{\Big{|}}_{T=T_{in}}-T_{ex}\,{\frac{\partial p_{ex}}{\partial T}}{\Big{|}}_{T=T_{ex}}\bigg)\,
 \Bigg] \, R^{2}=-\Big({p_{ex}}^{eff}-{p_{in}}^{eff}\Big)\, R^{2}
\label{efep}
\end{equation}
In the case of zero temperature, we have $F =-(p_{ex}-p_{in})  \, R^{2}$ as found in Ref. \cite{rrp}.
Looking at (\ref{rrp}),  we see that the left side contains the ``acceleration'' whereas the right side contains the ``force''.
When the pressure outside the bubble is  larger  than inside, i.e.  ${p_{ex}}^{eff}>{p_{in}}^{eff}$, the function $F$ will have negative values,
implying a force towards the center of the bubble, which leads to its collapse.  On the other hand if ${p_{ex}}^{eff}<{p_{in}}^{eff}$, the bubble will expand.
We could add a term in (\ref{rrp}) with the surface tension. However we
postpone the introduction of this term  because we do not know its precise form in the relativistic formulation of the RP equation and also because we expect
it to produce small effects compared to the uncertainties in the equations of state.

In the next subsection, we will obtain the non-relativistic limit of (\ref{rrp}) and will solve it analytically, confirming that when
${p_{ex}}^{eff}>{p_{in}}^{eff}$ (and ${p_{ex}}>{p_{in}}$ at zero temperature),  the bubble shrinks.

\subsection{Non-relativistic limit}

The non-relativistic limit for the RRP equation (\ref{rrp}) is obtained by expanding the Lorentz factors
around small values of $\dot{R}^{\,2}$.
Using this expansion and recalling that $x=r/R$  we find:
\begin{equation}
I_{1} \cong {\frac{1}{R^{5}}}\,\int_{0}^{R} d x \, x^{4} \,\Big(\varepsilon_{in} +{p_{in}}^{eff} \Big)
+{\frac{\dot{R}^{\,2}}{R^{7}}}\,\int_{0}^{R} d x \, x^{6} \,\Big(\varepsilon_{in} +  {p_{in}}^{eff} \Big)
\label{i1a}
\end{equation}
\begin{equation}
I_{2} \cong R\,\int_{R}^{\infty}{\frac{d x}{x^{2}}} \,\Big(\varepsilon_{ex} +
{p_{ex}}^{eff} \Big)+ R^{5}\,\dot{R}^{\,2}\int_{R}^{\infty}{\frac{d x}{x^{6}}} \,\Big(\varepsilon_{ex} +  {p_{ex}}^{eff}  \Big)
\label{i2a}
\end{equation}
The effective pressure and energy density of the gas and fluid phases are assumed to be constant in the  radial coordinate
and hence (\ref{i1a}) and (\ref{i2a}) can be simplified.
Substituting (\ref{efep}), (\ref{i1a}) and (\ref{i2a}) into (\ref{rrp}) and keeping only the linear terms in $\dot{R}^{\,2}$ we find:
$$
\bigg[{\frac{4}{5}}\,\Big(\varepsilon_{in} +{p_{in}}^{eff} \Big)+
\Big(\varepsilon_{ex} +{p_{ex}}^{eff} \Big)\bigg]\dot{R}^{\,2}
$$
\begin{equation}
+\bigg[{\frac{1}{5}}\,\Big(\varepsilon_{in} +{p_{in}}^{eff} \Big)
+\Big(\varepsilon_{ex} + {p_{ex}}^{eff} \Big)\bigg]R\,\ddot{R}
=-\Big({p_{ex}}^{eff}-{p_{in}}^{eff}\Big)
\label{rrpnr}
\end{equation}
which is the non-relativistic limit of the relativistic equation (\ref{rrp}).
We can rewrite the above equation in a compact form as:
\begin{equation}
R\,\ddot{R}+\alpha\,\dot{R}^{\,2}=\beta
\label{rrpnrf}
\end{equation}
where
\begin{equation}
\alpha \equiv {\frac{{\frac{4}{5}}\,\Big(\varepsilon_{in} +{p_{in}}^{eff} \Big)+
\Big(\varepsilon_{ex} +{p_{ex}}^{eff} \Big)}{{\frac{1}{5}}\,\Big(\varepsilon_{in} + {p_{in}}^{eff} \Big)
+\Big(\varepsilon_{ex} +{p_{ex}}^{eff} \Big)}}
\hspace{0.7cm} \textrm{and} \hspace{0.7cm}
\beta \equiv {\frac{-\Big( {p_{ex}}^{eff}- {p_{in}}^{eff}\Big)}{{\frac{1}{5}}\,\Big(\varepsilon_{in} + {p_{in}}^{eff} \Big)
+\Big(\varepsilon_{ex} +{p_{ex}}^{eff} \Big)}}
\label{coefsnrel}
\end{equation}
Using  the Sundman transformation method as described in \cite{nikolai}, we can find the
analytical solution of (\ref{rrpnrf}), which  is given by:
$$
t(R)=\Bigg({\frac{-2\mathcal{C}\alpha}{\beta}}\Bigg)^{1+{\frac{1}{2\alpha}}} \Bigg\{
\sqrt{{\frac{1}{2\mathcal{C}\alpha^{2}\,R^{2\alpha}}}+{\frac{\beta}{4\mathcal{C}^{2}\,\alpha^{3}}}}
\,\,\, _2F_1\Bigg(1+{\frac{1}{2\alpha}},{\frac{1}{2}};{\frac{3}{2}};{\frac{4\mathcal{C}^{2}\,\alpha^{3}}{\beta}}
\Big[{\frac{1}{2\mathcal{C}\alpha^{2}\,R^{2\alpha}}}+{\frac{\beta}{4\mathcal{C}^{2}\,\alpha^{3}}}\Big]\Bigg)
$$
\begin{equation}
+ \, \mathcal{A} \,\,\, _2F_1\Bigg(1+{\frac{1}{2\alpha}},{\frac{1}{2}};{\frac{3}{2}};
{\frac{4\mathcal{C}^{2}\,\alpha^{3}\,{\mathcal{A}}^{2}}{\beta}} \Bigg) \, \Bigg\}
\label{rrpnrwrtedansol}
\end{equation}
where the constants are:
\begin{equation}
\mathcal{C} = \Bigg({v_{0}}^{2}-{\frac{\beta}{\alpha}} \Bigg){\frac{{R_{0}}^{2\alpha}}{2}}
\hspace{1.0cm} \textrm{and} \hspace{1.0cm}
\mathcal{A} = {\frac{{R_{0}}^{1/4\alpha}}{2\mathcal{C}\alpha^{2}}}\, \sqrt{2\mathcal{C}\alpha^{2}+\alpha\,\beta\,{R_{0}}^{-1/2\alpha}}
\label{cconsts}
\end{equation}
The initial bubble radius is  $R_{0}=R(0)$   and  $v_{0}=\dot{R}(0)$ is the initial  velocity of the  bubble frontier.
The Hypergeometric functions in (\ref{rrpnrwrtedansol}) are given by \cite{hypergeom}:
\begin{equation}
_2F_1(a,b;c;z)={\frac{\Gamma(c)}{\Gamma(b)\,\Gamma(c-b)}}
\, \int_0^1 d\xi \, {\frac{{\xi}^{b-1} \,(1-\xi)^{c-b-1}}{(1-\xi \, z)^{a}}}
\label{hyperfunc}
\end{equation}
for $\Re(c)>\Re(b)>0$ . When the radius $R$ tends to zero, we have the collapse time $(t_{collapse})$ of the bubble.  Defining the following
variable in the first hypergeometric function in (\ref{rrpnrwrtedansol}):
\begin{equation}
w \equiv {\frac{4\mathcal{C}^{2}\,\alpha^{3}}{\beta}}
\Big[{\frac{1}{2\mathcal{C}\alpha^{2}\,R^{2\alpha}}}+{\frac{\beta}{4\mathcal{C}^{2}\,\alpha^{3}}}\Big]
\label{hyperfuncargfirst}
\end{equation}
we have  $w \rightarrow \infty$ when $R \rightarrow 0$ and in this regime the hypergeometric function is given by:
\begin{equation}
_2F_1\Bigg(1+{\frac{1}{2\alpha}},{\frac{1}{2}};{\frac{3}{2}};w\rightarrow \infty \Bigg) \cong -i{\frac{\sqrt{\pi}}{2}}\,
{\frac{\Gamma\Big({\frac{1}{2}}+{\frac{1}{2\alpha}}\Big)}{\Gamma\Big(1+{\frac{1}{2\alpha}}\Big)}}
\, {\frac{1}{\sqrt{w}}}
\label{hyperfuncargfirsta}
\end{equation}
Inserting (\ref{hyperfuncargfirsta}) into (\ref{rrpnrwrtedansol}) we have:
\begin{equation}
t_{collapse}=\Bigg({\frac{-2\mathcal{C}\alpha}{\beta}}\Bigg)^{1+{\frac{1}{2\alpha}}} \Bigg\{
{\frac{\sqrt{\pi}}{2}}\,\sqrt{{\frac{-\beta}{4\mathcal{C}^{2}\,\alpha^{3}}}}
\,\,{\frac{\Gamma\Big({\frac{1}{2}}+{\frac{1}{2\alpha}}\Big)}{\Gamma\Big(1+{\frac{1}{2\alpha}}\Big)}}
+\mathcal{A} \,\,\, _2F_1\Bigg(1+{\frac{1}{2\alpha}},{\frac{1}{2}};{\frac{3}{2}};
{\frac{4\mathcal{C}^{2}\,\alpha^{3}\,{\mathcal{A}}^{2}}{\beta}} \Bigg) \, \Bigg\}
\label{rrpnrwrtedansolcoll}
\end{equation}

We notice in (\ref{rrpnrwrtedansol}) and clearly in (\ref{rrpnrwrtedansolcoll}) that $\beta$ must be negative for the bubble collapse condition to hold.
This means that from (\ref{coefsnrel}) we must have ${p_{ex}}^{eff}>{p_{in}}^{eff}$. In the  zero temperature case we must have ${p_{ex}}>{p_{in}}$.

\section{Equations of state }

The temperature of the hadronic matter formed in the early stage of the nuclear collisions to be performed at FAIR and NICA  can be in the range of tens of MeV
going up to $\simeq 100 - 150$ MeV. In the following calculations we will discuss two extreme case, one with $T=0$ and another with temperatures around
$100 < T  < 140 $ MeV.

\subsection{Zero temperature}

In the literature there are numerous equations of state for both phases.
Here we shall employ two of them in the zero temperature case which are very well known and another one (mQCD),
which is very stiff and, when applied to the study of compact stars  is able to generate quark stars with two solar masses.

\subsubsection{MIT Bag Model at zero temperature}

The MIT bag model EOS for the QGP with quarks (equal masses $m_{q}$)  has the following pressure \cite{mit}:
\begin{equation}
{p_{MIT}}=\sum_{q=u}^{d,s}\,{\frac{\gamma_{q}}{6\pi^{2}}}\int_{0}^{k_{F}} dk \,{\frac{{k}^{4}}{\sqrt{m_{q}^{2}+k^{2}}}}\, -\mathcal{B}
\label{pressmitz}
\end{equation}
and energy density:
\begin{equation}
{\varepsilon_{MIT}}=\sum_{q=u}^{d,s}\,{\frac{\gamma_{q}}{2\pi^{2}}} \int_{0}^{k_{F}} dk \, k^{2}\, \sqrt{m_{q}^{2}+k^{2}}\, +\mathcal{B}
\label{energmitz}
\end{equation}
where $\mathcal{B}$ is the bag constant, $m_q =10$ MeV and
$\gamma_q=2\textrm{(spins)}\times 3
\textrm{(colors)}=6$ is the statistical factor for each quark flavor.  The quark density is given by \cite{mit}:
\begin{equation}
\rho_{MIT}=\sum_{q=u}^{d,s}\,{\frac{\gamma_{q}}{2\pi^{2}}} \int_{0}^{k_{F}} dk \, k^{2}
\label{quarkdensmitz}
\end{equation}
which defines the highest occupied level $k_{F}$.  Recalling that the baryon density is $\rho_{B}={\frac{\rho_{MIT}}{3}}$ we have from (\ref{quarkdensmitz})
the following relation:
\begin{equation}
\rho_{B}={\frac{{k_{F}}^{3}}{\pi^{2}}}
\label{barkf}
\end{equation}
which defines (\ref{pressmitz}) and (\ref{energmitz}) as functions of the baryon density $p_{MIT} =  p_{MIT}(\rho_{B})$ and
$\varepsilon_{MIT} =  \varepsilon_{MIT}(\rho_{B})$.

\subsubsection{Nonlinear Walecka Model at zero temperature}

The energy density and pressure of the Nonlinear Walecka Model (NLWM) are given respectively by \cite{serot,furn,we13}:
$$
{\varepsilon_{W}}(\rho_{B})={\frac{{g_{V}}^{2}}{2{m_{V}}^{2}}}{\rho_{B}}^{2}+
{\frac{{m_{S}}^{2}}{2{g_{S}}^{2}}}(M-M^{*})^{2}+
b{\frac{(M-M^{*})^{3}}{3{g_{S}}^{3}}}+
c{\frac{(M-M^{*})^{4}}{4{g_{S}}^{4}}}
$$
\begin{equation}
+{\frac{\gamma_s}{(2\pi)^{3}}}\int_{0}^{k_{F}} d^3{k}\hspace{0.2cm}\sqrt{{\vec{k}}^{2}+{M^{*}}^{2}}
\label{epswtz}
\end{equation}
and
$$
{p_{W}}(\rho_{B})={\frac{{g_{V}}^{2}}{2{m_{V}}^{2}}}{\rho_{B}}^{2}
-{\frac{{m_{S}}^{2}}{2{g_{S}}^{2}}}(M-M^{*})^{2}
-b{\frac{(M-M^{*})^{3}}{3{g_{S}}^{3}}}
-c{\frac{(M-M^{*})^{4}}{4{g_{S}}^{4}}}
$$
\begin{equation}
+{\frac{\gamma_s}{3(2\pi)^{3}}}\int_{0}^{k_{F}} d^3{k}\hspace{0.2cm}
{\frac{{\vec{k}}^{2}}{\sqrt{{\vec{k}}^{2}+{M^{*}}^{2}}}}
\label{preswtz}
\end{equation}
where the baryon density is given by:
\begin{equation}
\rho_{B}={\frac{\gamma_s}{6 \pi^{2}}}{k_{F}}^{3}
\label{bardwtz}
\end{equation}
and the nucleon effective mass by:
$$
{M^{*}}(\rho_{B})=M-{\frac{{g_{S}}^{2}}{{m_{S}}^{2}}}{\frac{\gamma_{s}}{(2\pi)^{3}}}\int_{0}^{k_{F}} d^3{k}{\frac{M^{*}}{\sqrt{{\vec{k}}^{2}+{M^{*}}^{2}}}}
$$
\begin{equation}
+{\frac{{g_{S}}^{2}}{{m_{S}}^{2}}}\Bigg[{\frac{b}{{g_{S}}^{3}}}(M-M^{*})^{2}+
{\frac{c}{{g_{S}}^{4}}}(M-M^{*})^{3}\Bigg]
\label{efmasswtz}
\end{equation}
and $\gamma_s=4$ is the nucleon degeneracy factor. As usual, $m_S$ and $m_V$  are the masses of the scalar and vector mesons and  $g_S$ and $g_V$
are their coupling constants to the nucleon.

\subsubsection{Mean Field Theory of QCD}

The mean field theory of QCD (mQCD) EOS was developed in \cite{we11} and it was successfully applied to stellar structure calculations of compact stars \cite{we12}
and also to  nonlinear wave propagation in the QGP \cite{we11a}.  This EOS is essentially the MIT one supplemented with a term proportional to the square of
the baryon density, which is included both in the pressure and in the energy density, yielding:
\begin{equation}
{p_{mQCD}} \, (\rho_{B})=\bigg({\frac{27\,g^{2}}{16\,{m_{G}}^{2}}}\bigg) \ {\rho_{B}}^{2} +{p_{MIT}}(\rho_{B})
\label{pressmftqcd}
\end{equation}
and:
\begin{equation}
{\varepsilon_{mQCD}} \, (\rho_{B})=\bigg({\frac{27\,g^{2}}{16\,{m_{G}}^{2}}}\bigg) \ {\rho_{B}}^{2} +{\varepsilon_{MIT}}(\rho_{B})
\label{energmftqcd}
\end{equation}
where $g$ is the QCD  coupling constant and $m_{G}$ is the dynamical gluon mass \cite{we11}.
Setting $g=0$ or performing $m_{G}\rightarrow \infty$ we recover the MIT bag model EOS.
As before, the baryon density $\rho_{B}$ is given by (\ref{barkf}).

\subsection{Finite temperature}

In the finite temperature case, we shall consider the Lattice QCD for the QGP and the Nonlinear Walecka Model.

\bigskip

\subsubsection{Lattice QCD}

The recent parametrization of a lattice simulation of
$SU(3)$ QCD matter at finite temperature, with gluons and quarks ($u$, $d$ and $s$ with equal masses) has the trace anomaly at finite chemical potential given by \cite{ratti,fodor14}:
$$
{\frac{{\varepsilon}_{Latt}(T,\mu)-3\, {p}_{Latt}(T,\mu)}{T^{4}}}=T{\frac{\partial}{\partial T}}
\Bigg[{\frac{{p}_{Latt}(T,\mu)}{T^{4}}}\Bigg]+{\frac{\mu^{2}}{T^{2}}}\,\chi_{2}
$$
\begin{equation}
={\frac{{\varepsilon}_{Latt}(T,0)-3\, {p}_{Latt}(T,0)}{T^{4}}}
+{\frac{\mu^{2}}{2T}}\,{\frac{d \chi_{2}}{d T}}
\label{fodortrami}
\end{equation}
where the chemical potential contribution is given by the function \cite{ratti}:
\begin{equation}
\chi_{2} (T)= e^{-h_{3}/{\tau}-h_{4}/{\tau}^{2}} \, f_{3} \, \Big[tanh (f_{4}\, {\tau}+f_{5})+1\Big]
\label{chemit}
\end{equation}
At zero chemical potential the parametrization is given by \cite{ratti,fodor14}:
\begin{equation}
{\frac{{\varepsilon}_{Latt}(T,0)-3\, {p}_{Latt}(T,0)}{T^{4}}}=e^{-h_{1}/{\tau}-h_{2}/{\tau}^{2}} \, \Bigg[h_{0}+
{\frac{f_{0} \, \Big[tanh(f_{1}\, {\tau}+f_{2})+1\Big]}{1+g_{1} \, {\tau}+g_{2} \, {\tau}^{2}}}   \Bigg] \,\,\, .
\label{fodortrazero}
\end{equation}
In the last two expressions the variable $\tau$ is given by $\tau=T/T_{c}$, with $T_{c}=200 $ $MeV$ being the critical temperature.  The dimensionless parameters are \cite{fodor14}: $h_{0} = 0.1396$, $h_{1} = -0.1800$, $h_{2} = 0.0350$, $f_{0} = 1.05$,
$f_{1} = 6.39$, $f_{2} = -4.72$, $g_{1} = -0.92$ and
$g_{2} = 0.57$ .  From \cite{ratti}: $h_{3} = -0.5022$, $h_{4} = 0.5950$, $f_{3} = 0.1359$, $f_{4} = 6.3290$ and $f_{5} = -4.8303$ .
The pressure is calculated from (\ref{fodortrami}):
\begin{equation}
{p}_{Latt}(T,\mu)=T^{4}\, \bigintss_{0}^{T} \, dT^{'} \,
{\frac{e^{-h_{1}/{\tau'}-h_{2}/{{\tau'}^{2}}}}{T^{'}}} \, \Bigg[h_{0}+
{\frac{f_{0} \, \Big[tanh(f_{1}\, {\tau'}+f_{2})+1\Big]}{1+g_{1} \, {\tau'}+g_{2} \, {{\tau'}^{2}}}}   \Bigg]
+{\frac{\chi_{2}}{2}}\, \mu^{2}  T^{2}
\label{prelatt}
\end{equation}
Inserting (\ref{prelatt}) into (\ref{fodortrami}) we find the following expression for the energy density \cite{we2015}:
$$
{\varepsilon}_{Latt}(T,\mu)=T^{4} \,e^{-h_{1}/{\tau}-h_{2}/{\tau}^{\,2}} \, \Bigg[h_{0}+
{\frac{f_{0} \, \Big[tanh(f_{1}\, {\tau}+f_{2})+1\Big]}{1+g_{1} \, {\tau}+g_{2} \, {\tau}^{2}}} \Bigg]
+{\frac{\mu^{2}}{2}}\, T^{3} \, {\frac{d \chi_{2}}{d T}}
$$
\begin{equation}
+3\,T^{4}\, \bigintss_{0}^{T} \, dT^{'} \,
{\frac{e^{-h_{1}/{\tau'}-h_{2}/{{\tau'}^{2}}}}{T^{'}}} \, \Bigg[h_{0}+
{\frac{f_{0} \, \Big[tanh(f_{1}\, {\tau'}+f_{2})+1\Big]}{1+g_{1} \, {\tau'}+g_{2} \, {{\tau'}^{2}}}}\Bigg]
+{\frac{3 \,\chi_{2}}{2 }} \, \mu^{2} T^{2}
\label{epslatt}
\end{equation}

\subsubsection{Nonlinear Walecka Model at finite temperature}

The equation of state of the nonlinear Walecka Model (NLWM) at finite temperature was  developed in \cite{serot,furn}.
The energy density and pressure are given respectively by:
$$
\varepsilon_{W}(T,\nu)={\frac{{g_{V}}^{2}}{2{m_{V}}^{2}}}{\rho_{B}}^{2}+
{\frac{{m_{S}}^{2}}{2{g_{S}}^{2}}}(M-M^{*})^{2}+
b{\frac{(M-M^{*})^{3}}{3{g_{S}}^{3}}}+
c{\frac{(M-M^{*})^{4}}{4{g_{S}}^{4}}}
$$
\begin{equation}
+{\frac{\gamma_s}{(2\pi)^{3}}}\int \, d^3{k}\hspace{0.2cm}\sqrt{{\vec{k}}^{2}+{M^{*}}^{2}}\Big(n+\bar{n}\Big)
\label{epswt}
\end{equation}
$$
p_{W}(T,\nu)={\frac{{g_{V}}^{2}}{2{m_{V}}^{2}}}{\rho_{B}}^{2}
-{\frac{{m_{S}}^{2}}{2{g_{S}}^{2}}}(M-M^{*})^{2}
-b{\frac{(M-M^{*})^{3}}{3{g_{S}}^{3}}}
-c{\frac{(M-M^{*})^{4}}{4{g_{S}}^{4}}}
$$
\begin{equation}
-T{\frac{\gamma_s}{(2\pi)^{3}}}\int \, d^3{k}\hspace{0.2cm}\Big[ ln\Big(1-n\Big)+ln\Big(1-\bar{n}\Big) \Big]
\label{preswt}
\end{equation}
The baryon density is given by:
\begin{equation}
\rho_{B}(T,\nu)= {\frac{\gamma_s}{(2\pi)^{3}}}\int \, d^3{k}\hspace{0.2cm} \Big(n-\bar{n}\Big)
\label{bardwt}
\end{equation}
The particle and antiparticle occupation numbers are:
\begin{equation}
n \equiv{\frac{1}{1+e^{(\sqrt{{\vec{k}}^{2}+{M^{*}}^{2}}-\nu)/T}}}
\hspace{2.5cm} \textrm{and} \hspace{2.5cm}
\bar{n}\equiv{\frac{1}{1+e^{(\sqrt{{\vec{k}}^{2}+{M^{*}}^{2}}+\nu)/T}}}
\label{distscav}
\end{equation}
where $\nu$ is the chemical potential.  The  nucleon effective mass ($M^{*}$) is given by the self-consistency relation:
\begin{equation}
M^{*}(T,\nu)=M-{\frac{{g_{S}}^{2}}{{m_{S}}^{2}}}{\frac{\gamma_{s}}{(2\pi)^{3}}}\int d^3{k}{\frac{M^{*}\big(n+\bar{n}\big)}{\sqrt{{\vec{k}}^{2}+{M^{*}}^{2}}}}
+{\frac{{g_{S}}^{2}}{{m_{S}}^{2}}}\Bigg[{\frac{b}{{g_{S}}^{3}}}(M-M^{*})^{2}+
{\frac{c}{{g_{S}}^{4}}}(M-M^{*})^{3}\Bigg]
\label{efmasswt}
\end{equation}
The nucleon degeneracy factor is $\gamma_s=4$ and the masses and couplings are given by \cite{serot,furn}: $M=939 \, MeV$,  $m_{S}=550 \, MeV$, $m_{V}=783 \, MeV$, $g_{S}=8.81$, $g_{V}=9.197$, $b = 13.47 \, fm^{-1}$ and $c = 43.127$.

We consider the case where the chemical potential is larger than the
temperature, in this case we can calculate the integrals in (\ref{epswt}), (\ref{preswt}) , (\ref{bardwt}) and (\ref{efmasswt}) in a simple way, following the approximations performed in \cite{tooper}.  The results are:
$$
\varepsilon_{W}(T,\nu)={\frac{{g_{V}}^{2}}{2{m_{V}}^{2}}}{\rho_{B}}^{2}+
{\frac{{m_{S}}^{2}}{2{g_{S}}^{2}}}(M-M^{*})^{2}+
b{\frac{(M-M^{*})^{3}}{3{g_{S}}^{3}}}+
c{\frac{(M-M^{*})^{4}}{4{g_{S}}^{4}}}
$$
$$
+{\frac{2 \,T^{2}\,{(M^{*})}^{2}}{\pi^{2}}}\sum_{n=0}^{\infty}{\frac{(-1)^{n}}{(n+1)^{2}}}
\, \Big[ e^{(n+1)\nu/T}+e^{-(n+1)\nu/T} \Big] \, \Bigg\{ 3K_{2}\Big[(n+1)M^{*}/T]
$$
\begin{equation}
+{\frac{(n+1)M^{*}}{T}}K_{1}\Big[(n+1)M^{*}/T] \Big] \, \Bigg\}
\label{epswtsol}
\end{equation}
$$
p_{W}(T,\nu)={\frac{{g_{V}}^{2}}{2{m_{V}}^{2}}}{\rho_{B}}^{2}
-{\frac{{m_{S}}^{2}}{2{g_{S}}^{2}}}(M-M^{*})^{2}
-b{\frac{(M-M^{*})^{3}}{3{g_{S}}^{3}}}
-c{\frac{(M-M^{*})^{4}}{4{g_{S}}^{4}}}
$$
\begin{equation}
-{\frac{2 \,T^{2}\,{(M^{*})}^{2}}{\pi^{2}}}\sum_{n=1}^{\infty}{\frac{(-1)^{n}}{n^{2}}}
\, \Big[ e^{n\nu/T}+e^{-n\nu/T} \Big] \, K_{2}\Big[nM^{*}/T \Big]
\label{preswtsol}
\end{equation}
\begin{equation}
\rho_{B}(T,\nu)= {\frac{2 \,T\,{(M^{*})}^{2}}{\pi^{2}}}\sum_{n=0}^{\infty}{\frac{(-1)^{n}}{(n+1)}}
\, \Big[e^{(n+1)\nu/T}-e^{-(n+1)\nu/T}\Big] \, K_{2}\Big[(n+1)M^{*}/T \Big]
\label{bardwtsol}
\end{equation}
and
$$
M^{*}(T,\nu)=M+{\frac{{g_{S}}^{2}}{{m_{S}}^{2}}}\,
{\frac{2 \,T\,{(M^{*})}^{2}}{\pi^{2}}}\sum_{n=1}^{\infty}{\frac{(-1)^{n}}{n}}
\, \Big[ e^{n\nu/T}+e^{-n\nu/T} \Big] \, K_{1}\Big[nM^{*}/T \Big]
$$
\begin{equation}
+{\frac{{g_{S}}^{2}}{{m_{S}}^{2}}}\Bigg[{\frac{b}{{g_{S}}^{3}}}(M-M^{*})^{2}+
{\frac{c}{{g_{S}}^{4}}}(M-M^{*})^{3}\Bigg]
\label{efmasswtsol}
\end{equation}
where $K_{1}$ and $K_{2}$ are the modified Bessel functions.

\section{Numerical results  and discussion}

The thermodynamical functions $\varepsilon$ and $p$ are
assumed to be independent of  the time and of the radial coordinate and hence the integrals in (\ref{prei1}) and (\ref{prei2})  can be simplified  as follows:
\begin{equation}
I_{1} = \Big(\varepsilon_{in} + {p_{in}}^{eff}   \Big) \int_{0}^{1} dx \, x^{4} \, {\gamma_{in}}^{2}
=\Big(\varepsilon_{in} + {p_{in}}^{eff}  \Big) \int_{0}^{1} dx \, {\frac{x^{4}}{(1-x^{2}\,\dot{R}^{\,2})}}
\label{fi1}
\end{equation}
and
\begin{equation}
I_{2} = \Big(\varepsilon_{ex} +{p_{ex}}^{eff}  \Big) \int_{1}^{\infty}{\frac{dx}{x^{2}}} \,{\gamma_{ex}}^{2}
=\Big(\varepsilon_{ex} + {p_{ex}}^{eff}  \Big)\int_{1}^{\infty} \, dx \, {\frac{x^{2}}{(x^{4}-\dot{R}^{\,2})}}
\label{fi2}
\end{equation}
These integrals can be written in terms of hypergeometric functions. Inserting (\ref{fi1}) and (\ref{fi2}) into  (\ref{rrp}) we obtain (after some algebra)
the  RRP  equation:
\begin{equation}
A \, R\,\dot{R}^{\,2}\,\ddot{R} \, + \, B  \,R\,\ddot{R} \, + \, C  \, \dot{R}^{\,2}  = -\Big(  {p_{ex}}^{eff} -  {p_{in}}^{eff} \Big)
\label{rrpf}
\end{equation}
where
\begin{equation}
A \, = \, \Bigg[\, {\frac{2}{7}} \, \Big(\varepsilon_{in} + {p_{in}}^{eff} \Big) \,
\,\, _2F_1\Bigg(2,{\frac{7}{2}};{\frac{9}{2}};\dot{R}^{\,2}\Bigg) \,
+ \, {\frac{2}{5}} \,  \Big(\varepsilon_{ex} + {p_{ex}}^{eff}  \Big) \,
\,\, _2F_1\Bigg(2,{\frac{5}{4}};{\frac{9}{4}};\dot{R}^{\,2}\Bigg)\Bigg]
\label{adef}
\end{equation}
\begin{equation}
B \, = \, \Bigg[ \, {\frac{1}{5}} \,  \Big(\varepsilon_{in} + {p_{in}}^{eff} \Big)\,
\,\, _2F_1\Bigg(1,{\frac{5}{2}};{\frac{7}{2}};\dot{R}^{\,2}\Bigg)
+\Big(\varepsilon_{ex} + {p_{ex}}^{eff} \Big)
\,\, _2F_1\Bigg(1,{\frac{1}{4}};{\frac{5}{4}};\dot{R}^{\,2}\Bigg) \Bigg]
\label{bdef}
\end{equation}
\begin{equation}
C \, = \, \Bigg[ \frac{4}{5} \, \Big(\varepsilon_{in} +{p_{in}}^{eff} \Big) \,
\,\, _2F_1\Bigg(1,{\frac{5}{2}};{\frac{7}{2}};\dot{R}^{\,2}\Bigg)
+\Big(\varepsilon_{ex} + {p_{ex}}^{eff} \Big)
\,\, _2F_1\Bigg(1,{\frac{1}{4}};{\frac{5}{4}};\dot{R}^{\,2}\Bigg) \Bigg]
\label{cdef}
\end{equation}

\subsection{Zero temperature}

In this subsection we present the results obtained at $T=0$. This is an extreme case. In the collisions relevant for this work the temperature is certainly bigger
than zero, although there are no definitive conclusions about its value. From the practical point of view, the thermodynamics of strongly interacting systems at $T=0$
and at $T= 10 - 40$ MeV is qualitatively not very different. For example, in the NLWM the effective mass $M^*$ is almost constant with $T$ in this region and $p(T)$ and
$\varepsilon(T)$ change slowly with $T$. In view of this, it is interesting to do calculations at $T=0$ because we can compare the results with those obtained in the
context of compact stars. In the case of higher temperatures $(T \simeq 100 \, \mbox{MeV})$ we use to the formalism at finite temperature. This will be done in the
next subsection.

\begin{figure}
\begin{center}
\subfigure[ ]{\label{fig2a}
\includegraphics[width=0.48\textwidth]{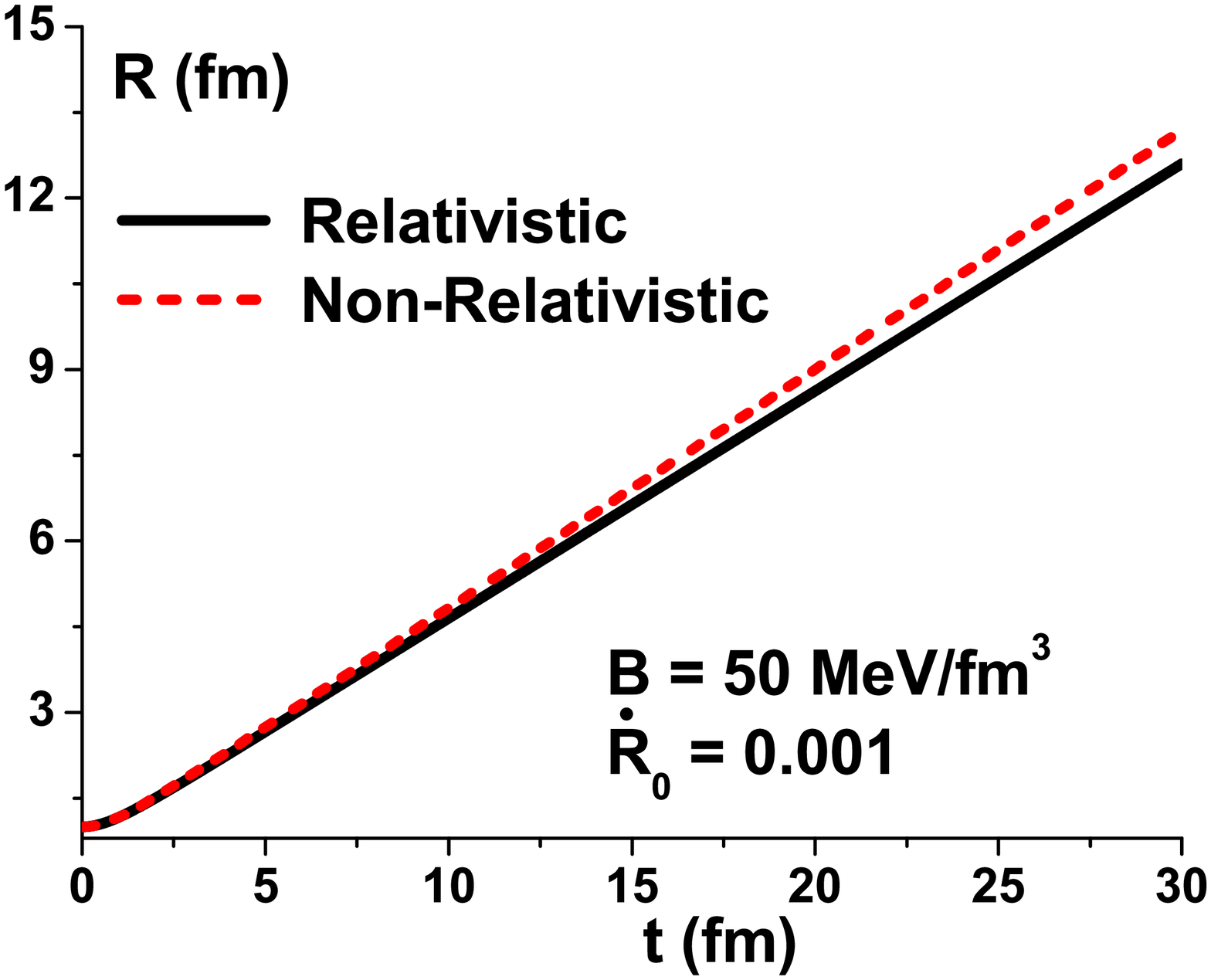}} 
\subfigure[ ]{\label{fig2b}
\includegraphics[width=0.48\textwidth]{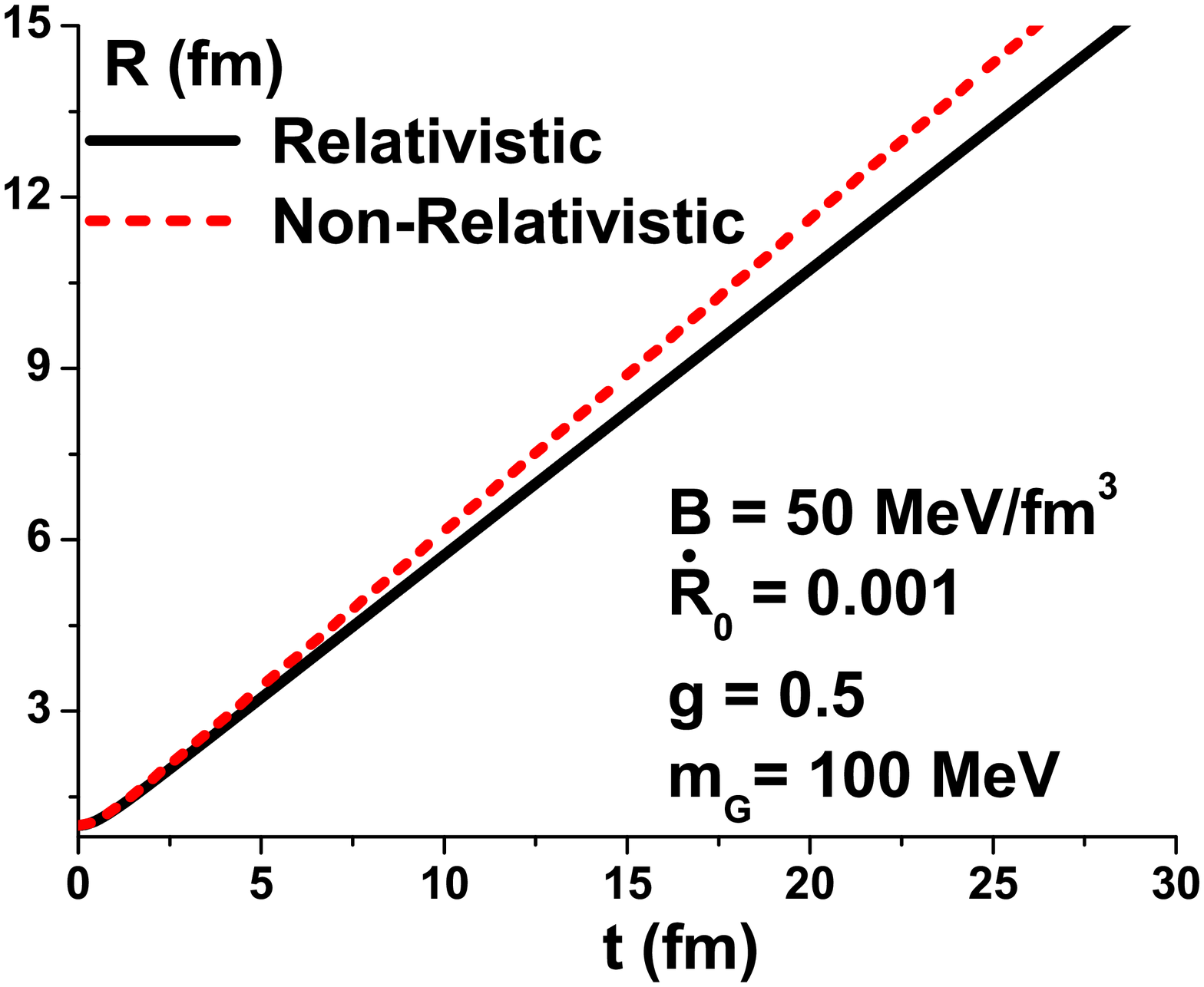}}
\subfigure[ ]{\label{fig2c}
\includegraphics[width=0.48\textwidth]{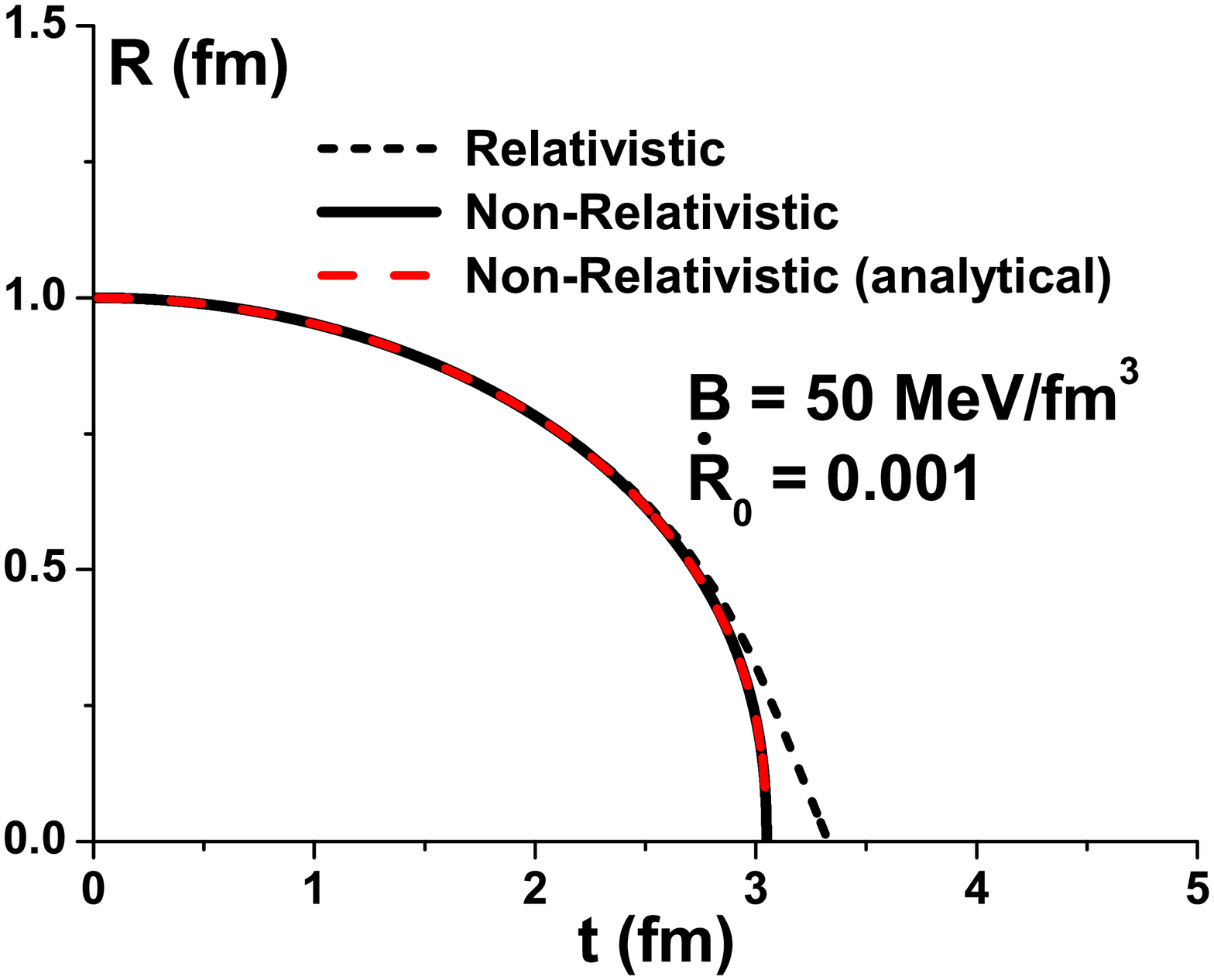}}
\subfigure[ ]{\label{fig2d}
\includegraphics[width=0.48\textwidth]{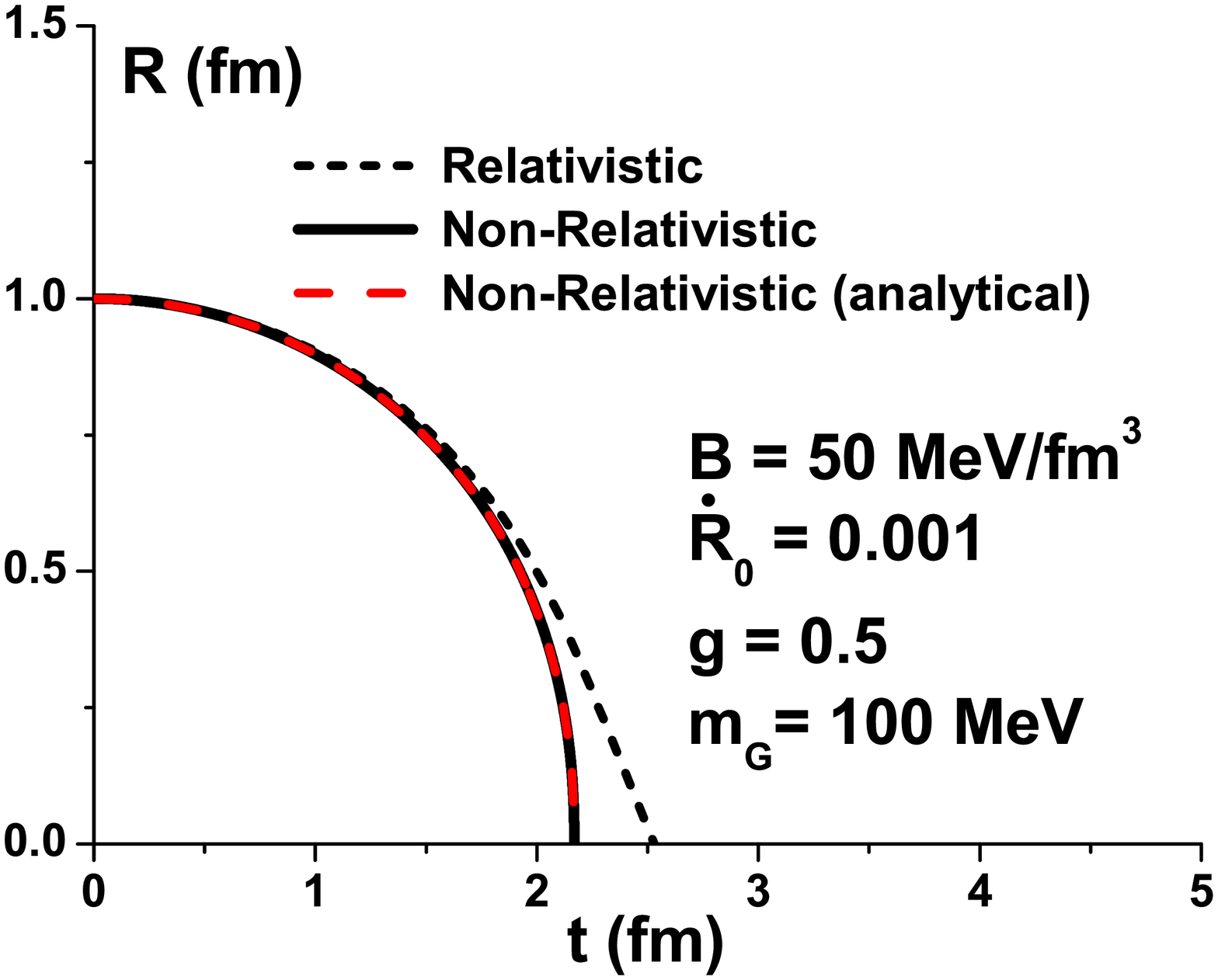}} 
\end{center}
\caption{Comparison between the solutions of the relativistic and non-relativistic RP equations in the high density phase ($\rho_B > \rho_c$).
a) and b): expanding trajectories of  a QGP bubble for different EOS. c) and d): collapsing trajectories of a hadron gas bubble for the same two EOS.}
\label{fig2}
\end{figure}

In order to estimate the importance of the relativistic effects,
in Fig. \ref{fig2}  we compare the solutions of Eq. (\ref{rrpf}) with the solutions of Eq. (\ref{rrpnrf}). The figure illustrates what happens
during the initial stage of  an intermediate energy nuclear collision. If the critical density is reached, bubbles of the quark phase start to grow (Figs. 2a and
2b) and later the  bubbles  of the hadron phase start to shrink (Figs. 2c and 2d).
From the figures we see that the  relativistic effects are very small. For completeness we also show (in thick  long dashed lines) in Figs. 2c and 2d
the analytical results obtained with (\ref{rrpnrwrtedansol}), rewritten as $R(t)$. As expected, there is a almost complete agreement between the numerical
solution of  (\ref{rrpnrf}) and the analytical solution.

In all cases the initial velocity of the surface is small and grows up to values which are still smaller
than  the typical values of the sound velocity. This result is welcome and does not contradict  the ``adiabatic'' assumption made before.  The only
significant, although still small, difference between the relativistic and non-relativistic cases arises at late times, when the pressure difference acted
long  enough to accelerate the bubble frontier to large velocities. The non-relativistic solutions do not have any velocity upper bound and  thus they exhibit
steeper slopes in the figures at late times. This same conclusion was reached in \cite{rrp} with a polytropic equation of state and for the cases  where
$p_{in} << p_{ex} $. From Fig. \ref{fig2} we can conclude that the collapse of a hadron gas bubble merged in a soft quark-gluon plasma occurs at  $3.3 \, fm$ in the
relativistic case and at  $3.0 \, fm$ in the    non-relativistic case for both numerical and analytical solutions.
Using a much stiffer equation of state for the QGP the collapse occurs sooner, at  $2.5 \, fm$ in the relativistic case at  $2.2 \, fm$ in the
non-relativistic case for both numerical and analytical solutions.

\begin{figure}
\begin{center}
\subfigure[ ]{\label{fig3a}
\includegraphics[width=0.48\textwidth]{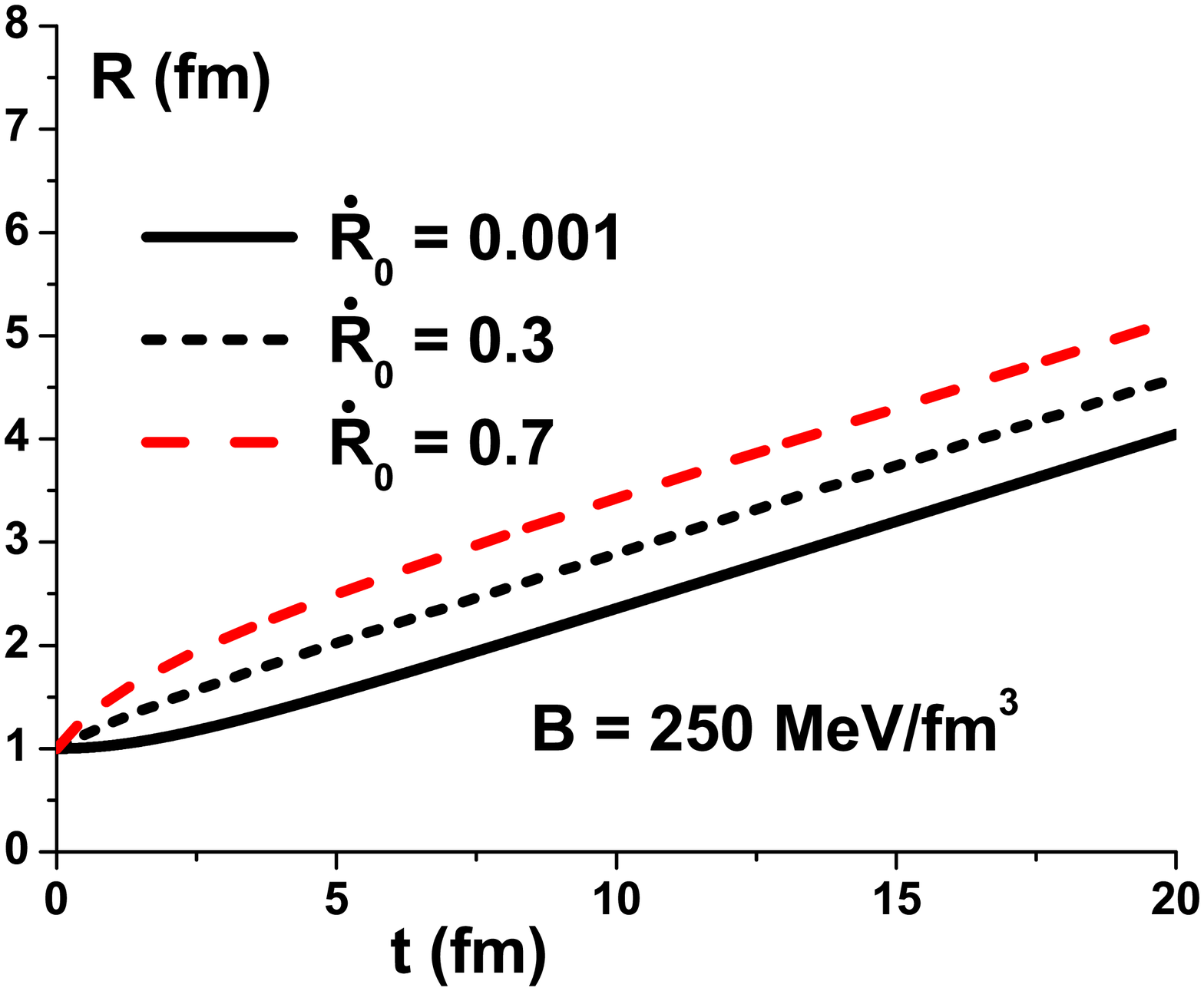}}
\subfigure[ ]{\label{fig3d}
\includegraphics[width=0.48\textwidth]{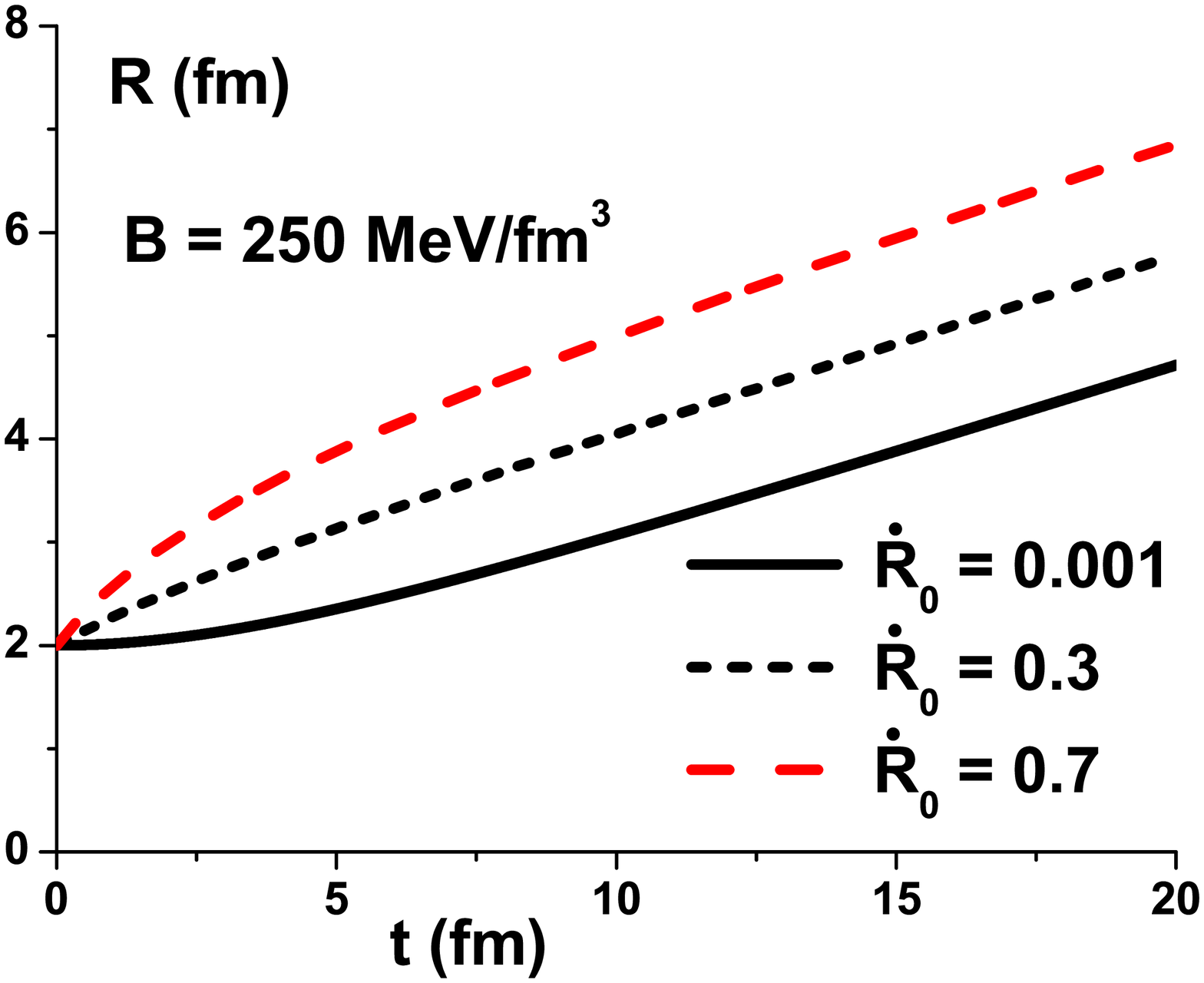}}
\subfigure[ ]{\label{fig3b}
\includegraphics[width=0.48\textwidth]{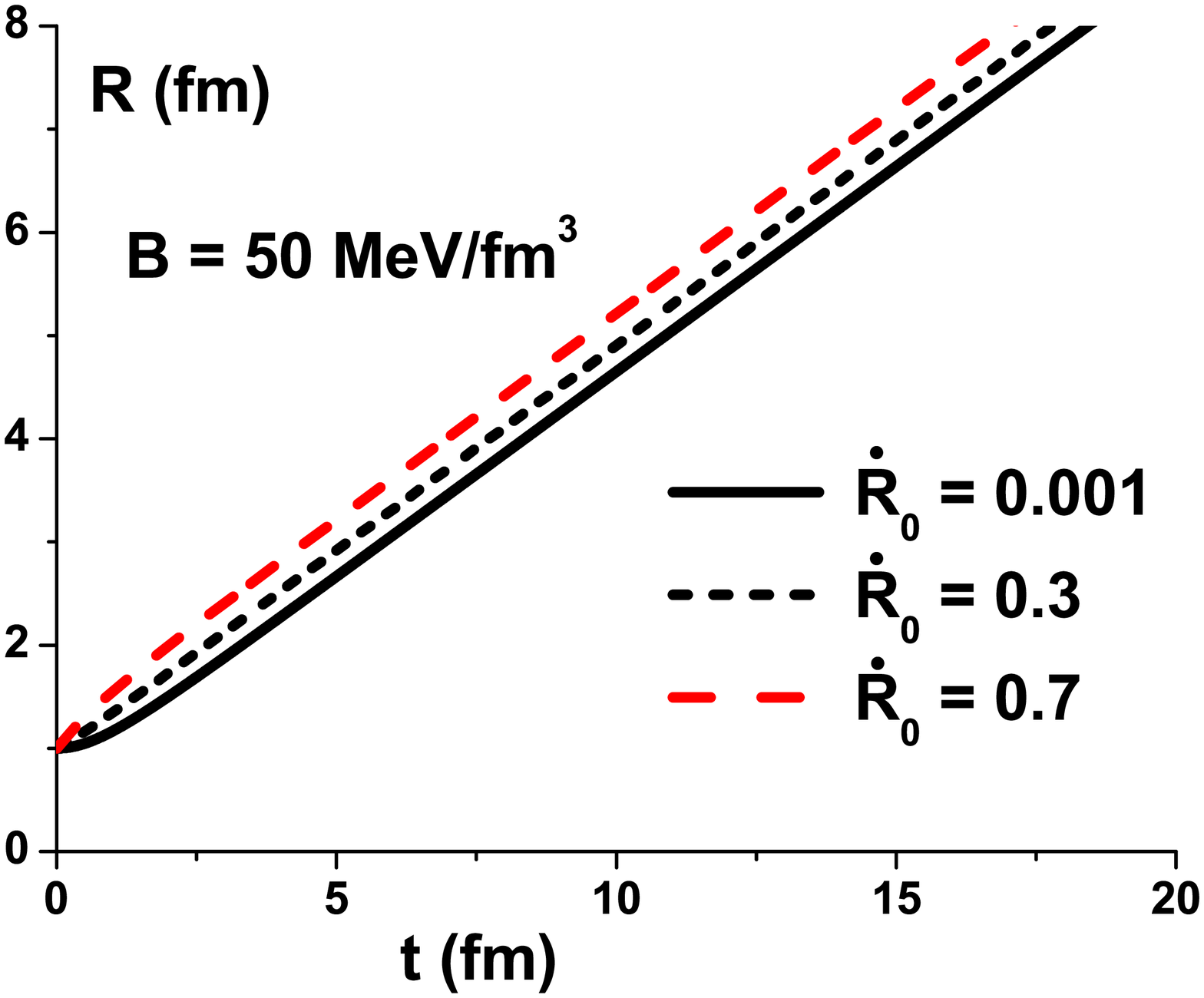}}
\subfigure[ ]{\label{fig3e}
\includegraphics[width=0.48\textwidth]{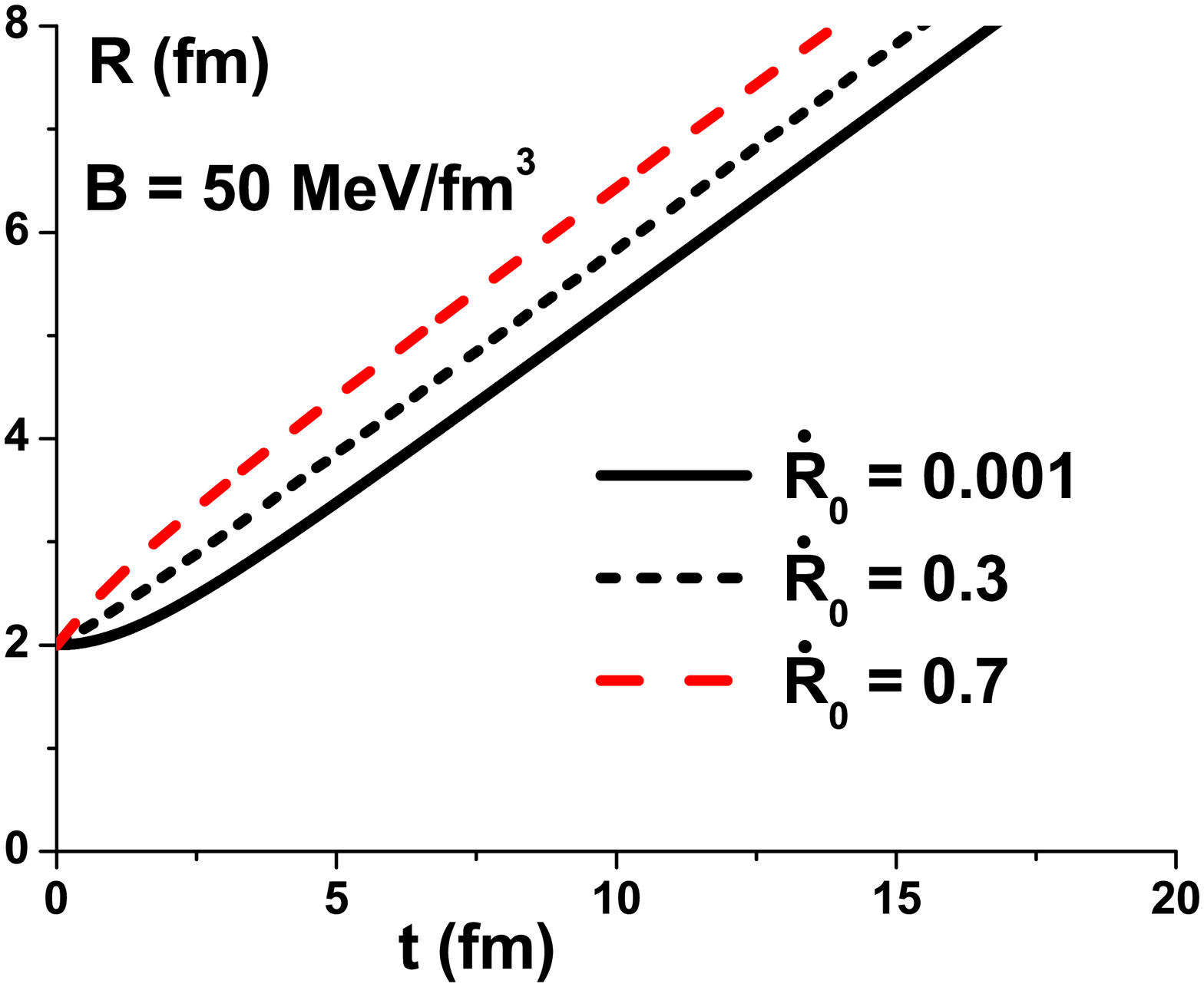}}
\subfigure[ ]{\label{fig3c}
\includegraphics[width=0.48\textwidth]{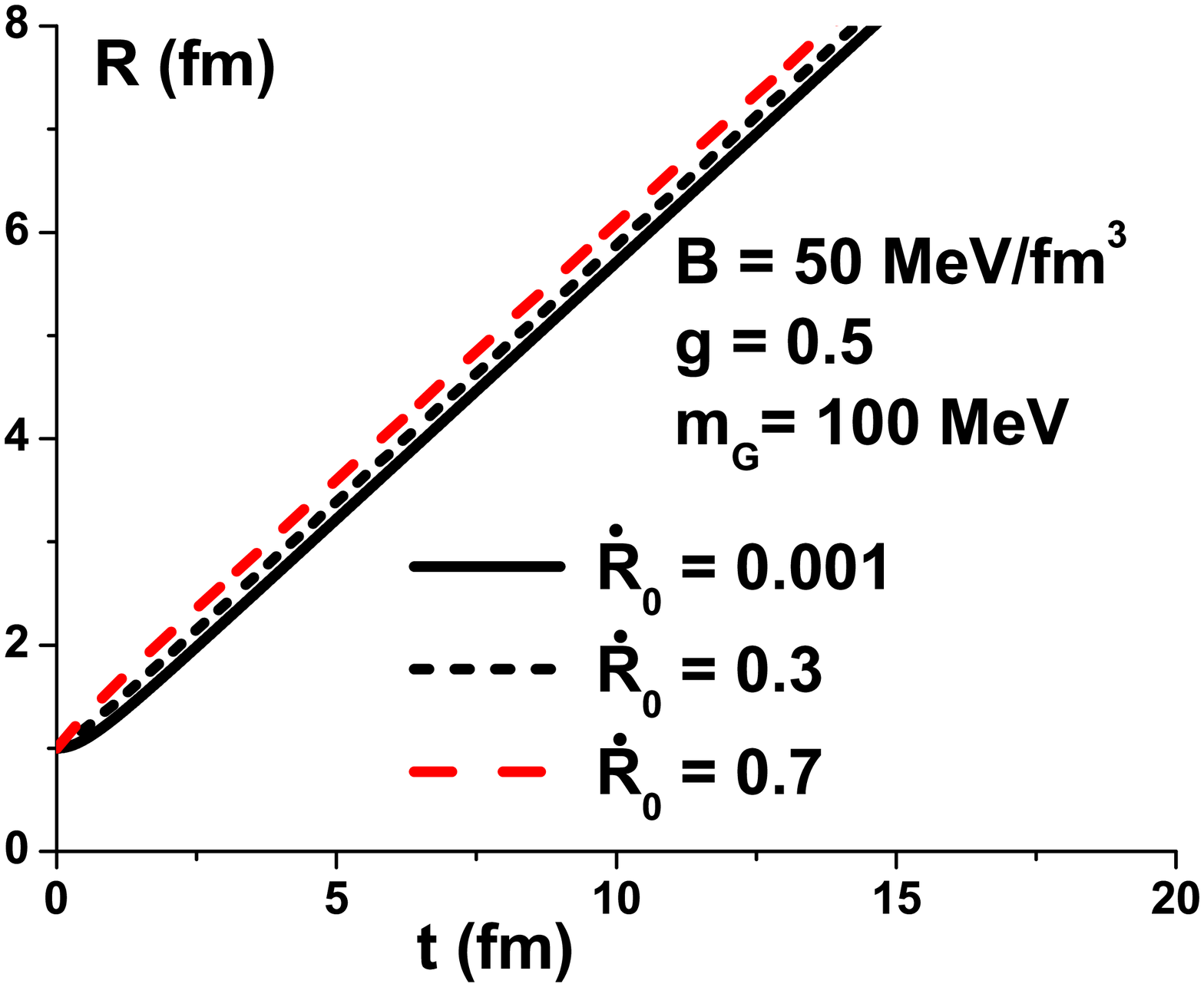}}
\subfigure[ ]{\label{fig3f}
\includegraphics[width=0.48\textwidth]{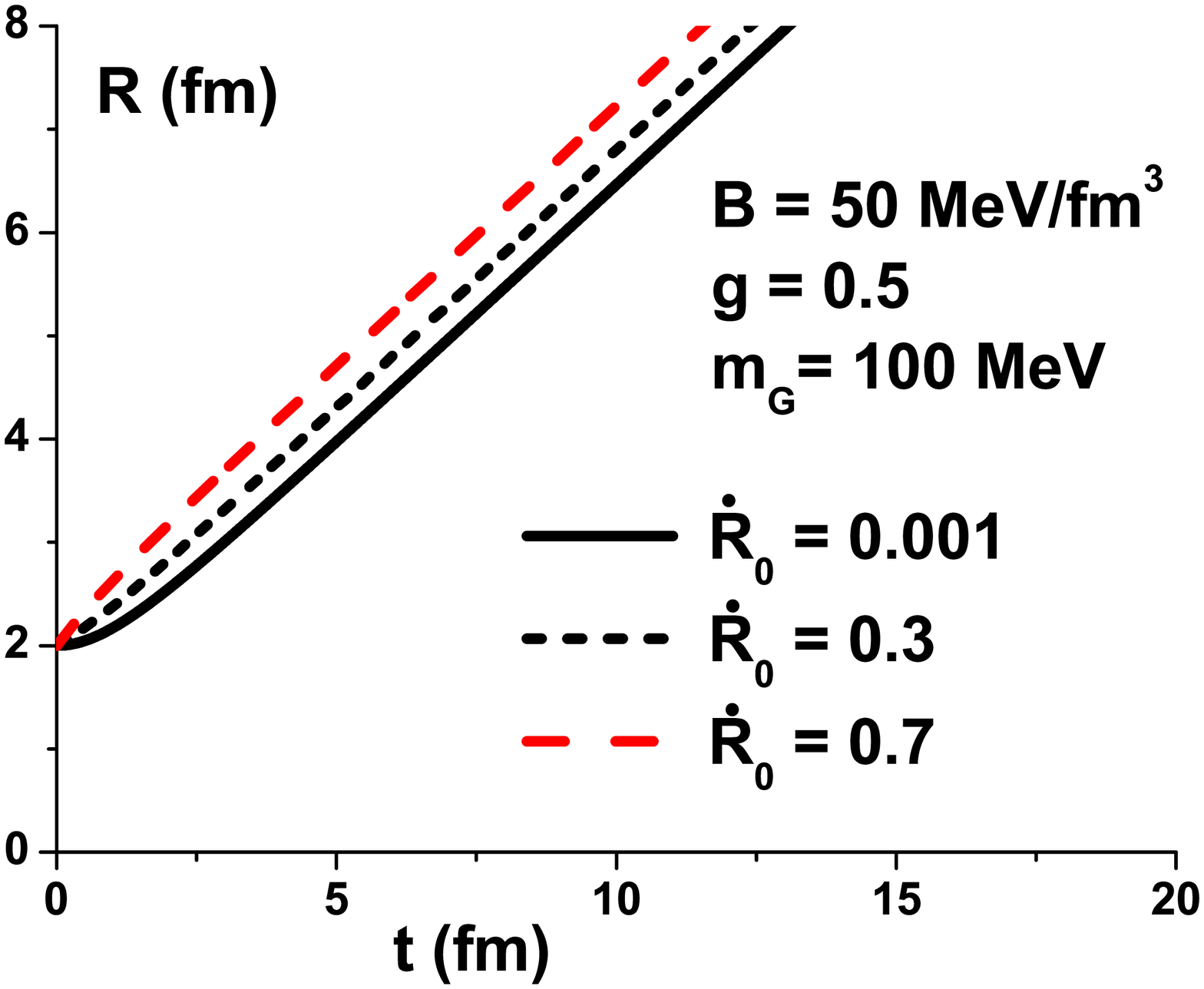}}
\end{center}
\caption{Time evolution of a QGP bubble immersed in a hadron gas: numerical solutions of (\ref{rrpf}) for several initial conditions.
In all situations: $\rho_{ex}=0.35\,fm^{-3}$ and $\rho_{in}=1.0\,fm^{-3}$ .}
\label{fig3}
\end{figure}

In Fig. \ref{fig3} we present the numerical solutions of equation (\ref{rrpf}) in the case of a QGP (with the MIT or mQCD equation of state) bubble in a
hadron external medium (with NLWM equation of state). We use the three equations of state mentioned in the previous section.

From the top to bottom the equations of state are harder and, as a consequence, the expansion is faster. Since the scales are the same we can observe this feature
just by looking at the increasing slopes of the curves. Changing the initial velocity of the bubble frontier ($\dot{R}(t=0)$) does not significantly change the trajectory
$R(t)$, especially for harder equations of state. Since Eq. (\ref{rrpf})  is a nonlinear equation we might expect some nontrivial dependence of the solutions on the
initial conditions.  To check this we have repeated the calculation changing the initial radius of the bubble from $1$ fm (the left panels in Fig. \ref{fig3}) to
$2$ fm (the right panels in \ref{fig3}). We observe that the initially larger bubbles expand somewhat slower than the smaller ones. However this difference is less
pronounced for harder equations of state.   Looking at these numbers we conclude that both the smaller and the larger
QGP bubbles  will take at least (see Fig. \ref{fig3e} and Fig. \ref{fig3f} with the hardest EOS)  $10$ fm to reach a typical size of the system $\simeq 6$ fm.
This is too long compared with $1-2$ fm, which, according to several simulations, is  the starting time of the QGP hydrodynamical evolution in heavy ion collisions
at RHIC and LHC. This time is also  large compared to the typical total collision times at FAIR, which are in the range $15-20$ fm.

\begin{figure}
\begin{center}
\subfigure[ ]{\label{fig4a}
\includegraphics[width=0.48\textwidth]{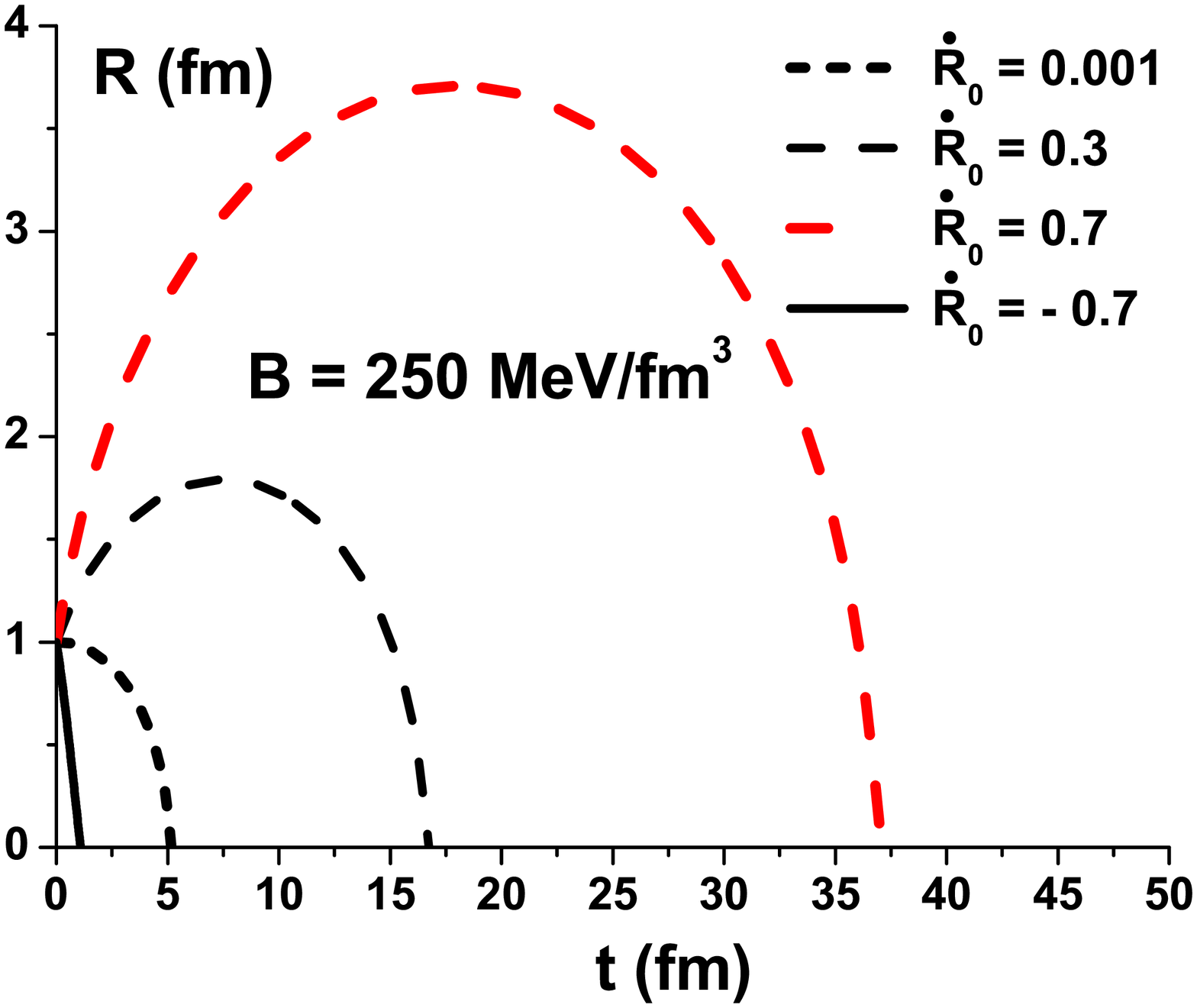}}
\subfigure[ ]{\label{fig4d}
\includegraphics[width=0.48\textwidth]{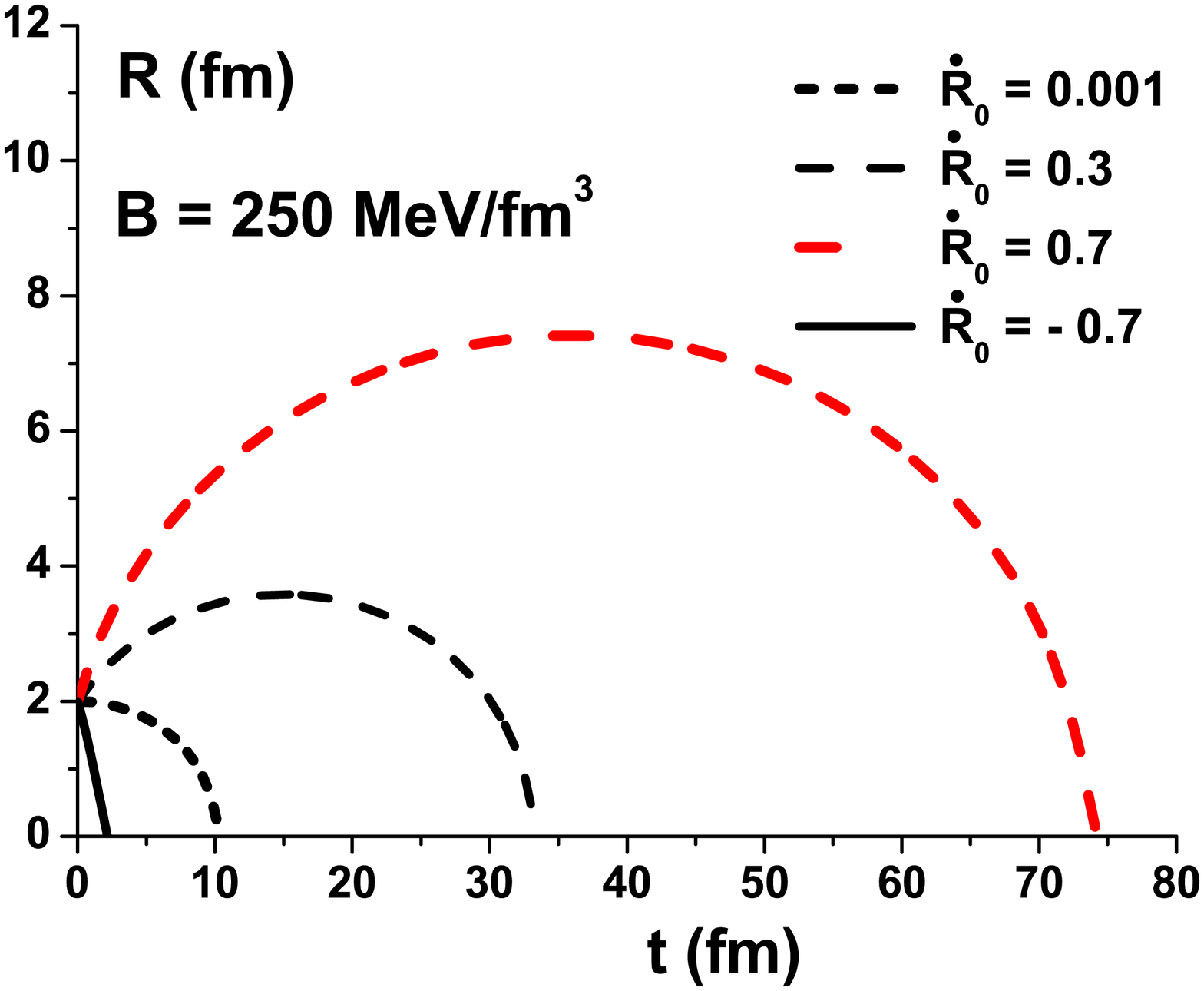}}
\subfigure[ ]{\label{fig4b}
\includegraphics[width=0.48\textwidth]{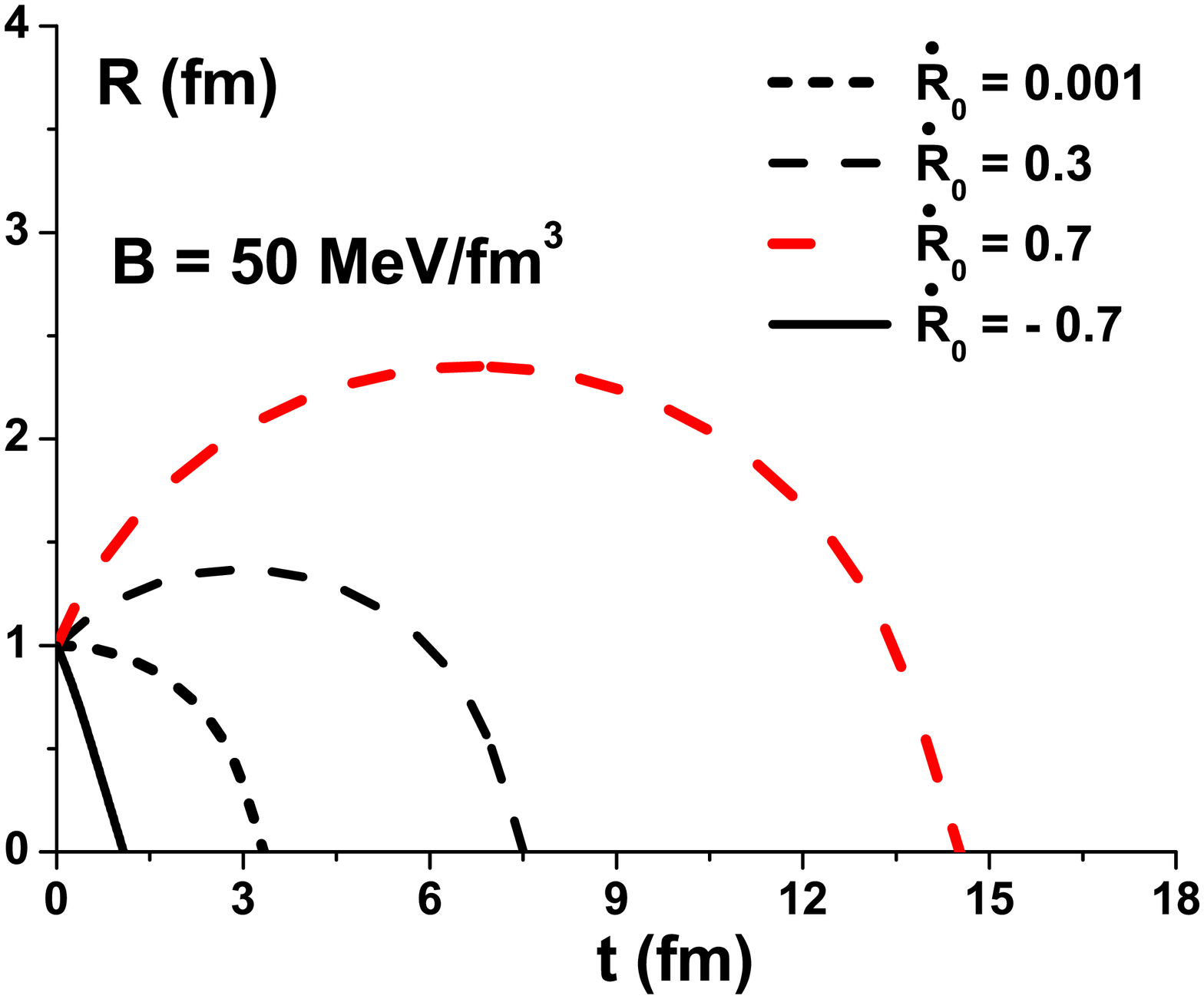}}
\subfigure[ ]{\label{fig4e}
\includegraphics[width=0.48\textwidth]{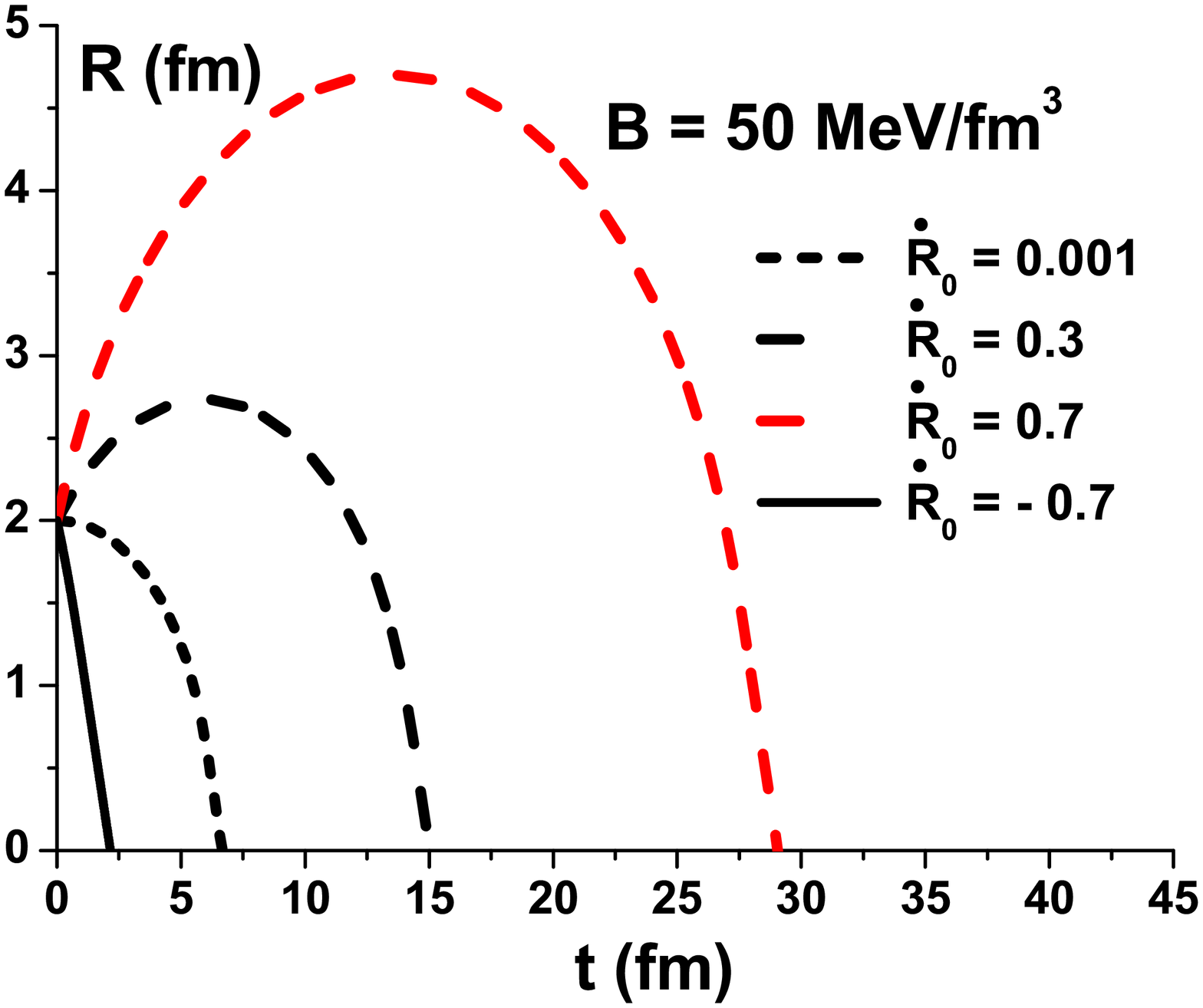}}
\subfigure[ ]{\label{fig4c}
\includegraphics[width=0.48\textwidth]{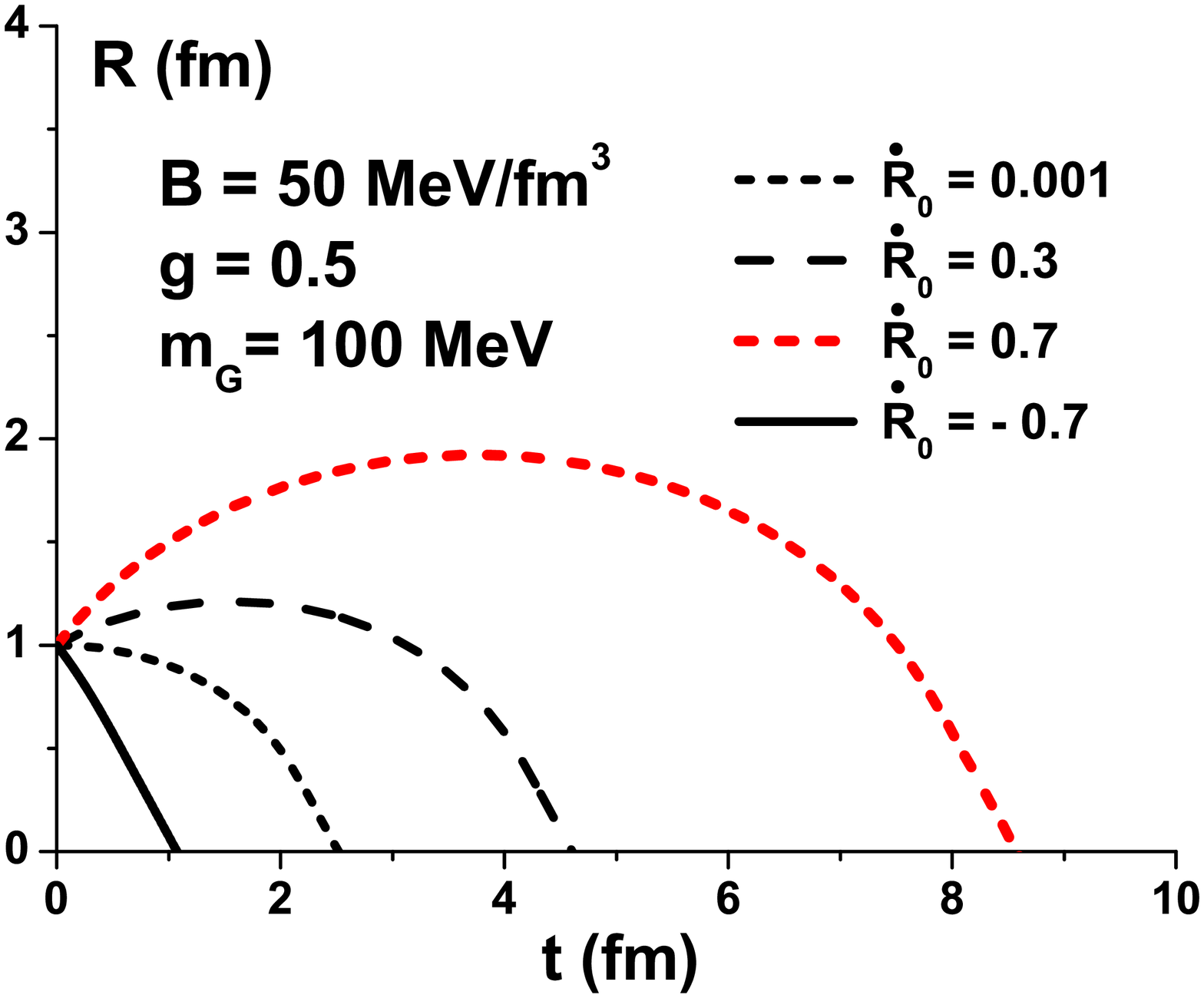}}
\subfigure[ ]{\label{fig4f}
\includegraphics[width=0.48\textwidth]{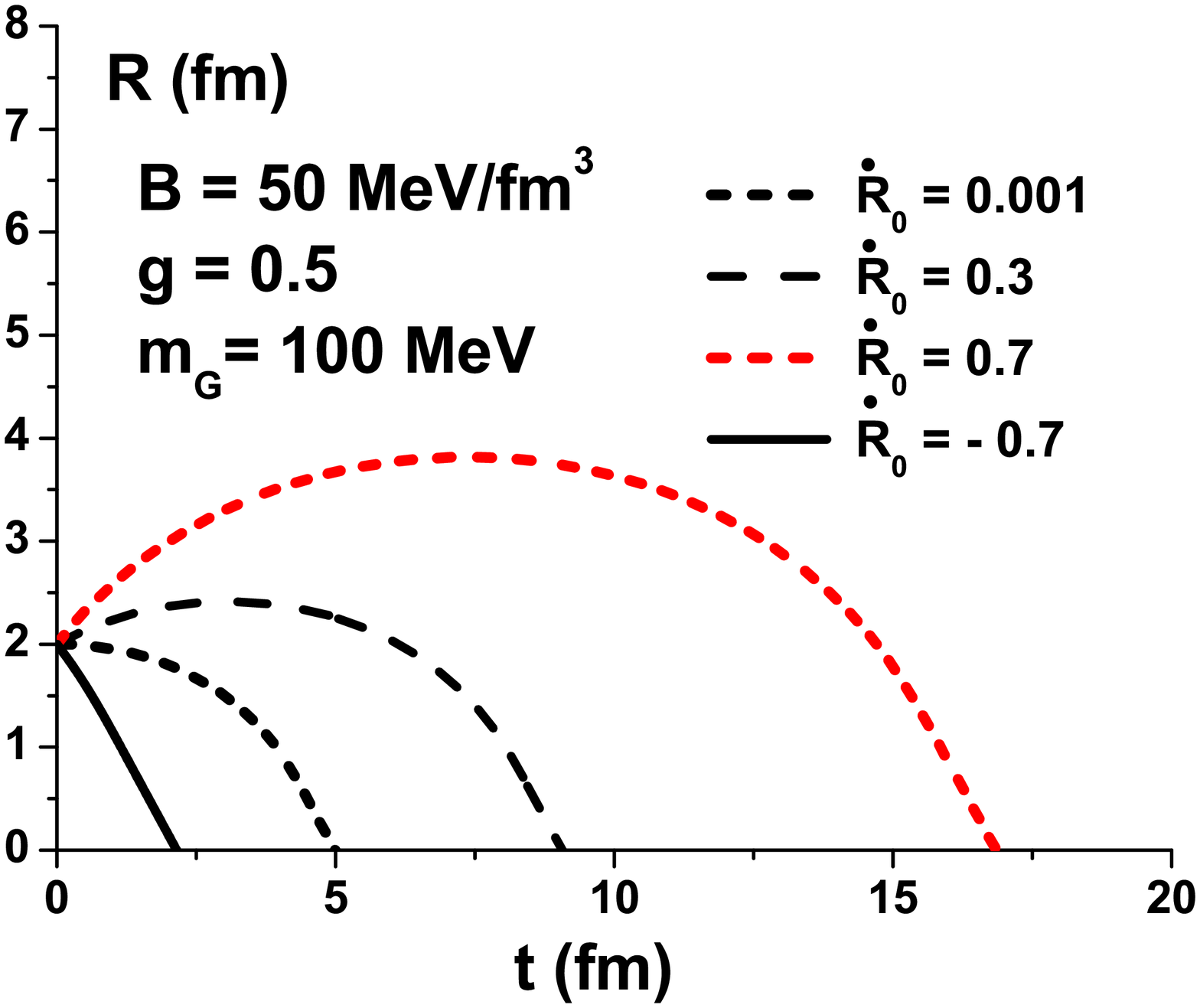}}
\end{center}
\caption{Time evolution of a hadron gas bubble immersed in QGP: numerical solutions of (\ref{rrpf}) for several initial conditions.
For (a) and (b): $\rho_{ex}=1.0\,fm^{-3}$ and $\rho_{in}=0.3\,fm^{-3}$ .
For (c) to (f): $\rho_{ex}=0.6\,fm^{-3}$ and $\rho_{in}=0.4\,fm^{-3}$ .}
\label{fig4}
\end{figure}

In Fig. \ref{fig4} we show the time evolution of a hadron gas bubble immersed in QGP given by the numerical solutions of the RRP
equation (\ref{rrpf}) with the  equations of state discussed in the previous section.  This figure is quite similar to Fig. \ref{fig3},
the only difference being which phase is ``inside''  and ``outside''.  Now, when we move from top to bottom (to harder QGP EOS), the hadron gas bubbles
shrink faster. They shrink even if the initial velocity is outwards.  As expected, the time of collapse is smaller for  harder QGP EOS. In any case it is
relatively large and it is approximately proportional to the initial size of the bubble.

While the trends observed in Figs. \ref{fig3}  and \ref{fig4}  were expected, now we have quantitative estimates for the relevant time scales of these
processes, which could be useful for nuclear collisions at FAIR. From the results in the figures we conclude that the time scale of the bubble motion is too
large compared to previous expectations based on high energy collisions. This suggests that this route to the quark gluon plasma, i.e. formation and expansion
of QGP bubbles, is not very effective. Either some extra ingredient is missing or the route is a different one.

\subsection{Finite temperature}

\begin{figure}
\begin{center}
\subfigure[ ]{\label{fig5a}
\includegraphics[width=0.48\textwidth]{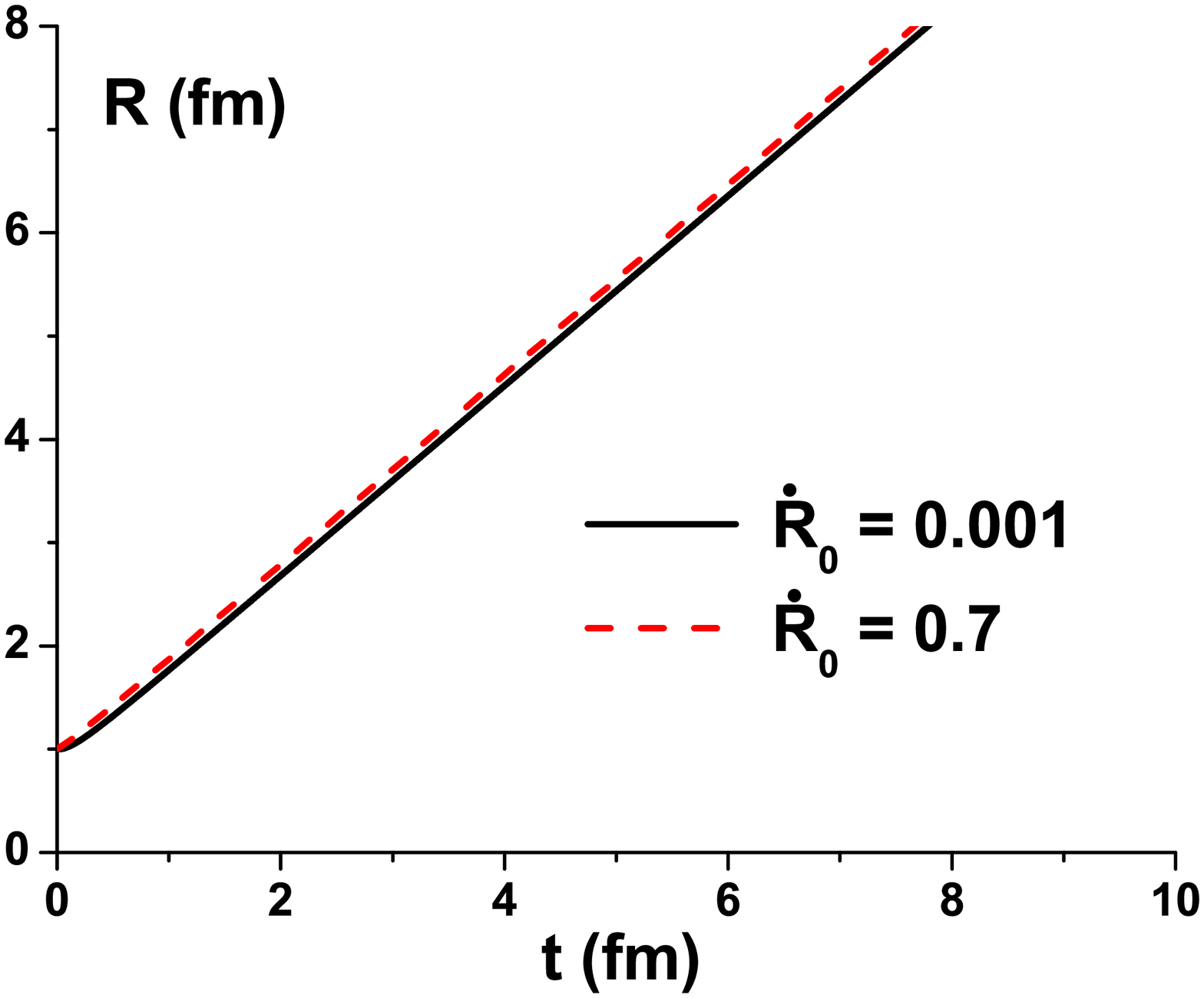}}
\subfigure[ ]{\label{fig5c}
\includegraphics[width=0.48\textwidth]{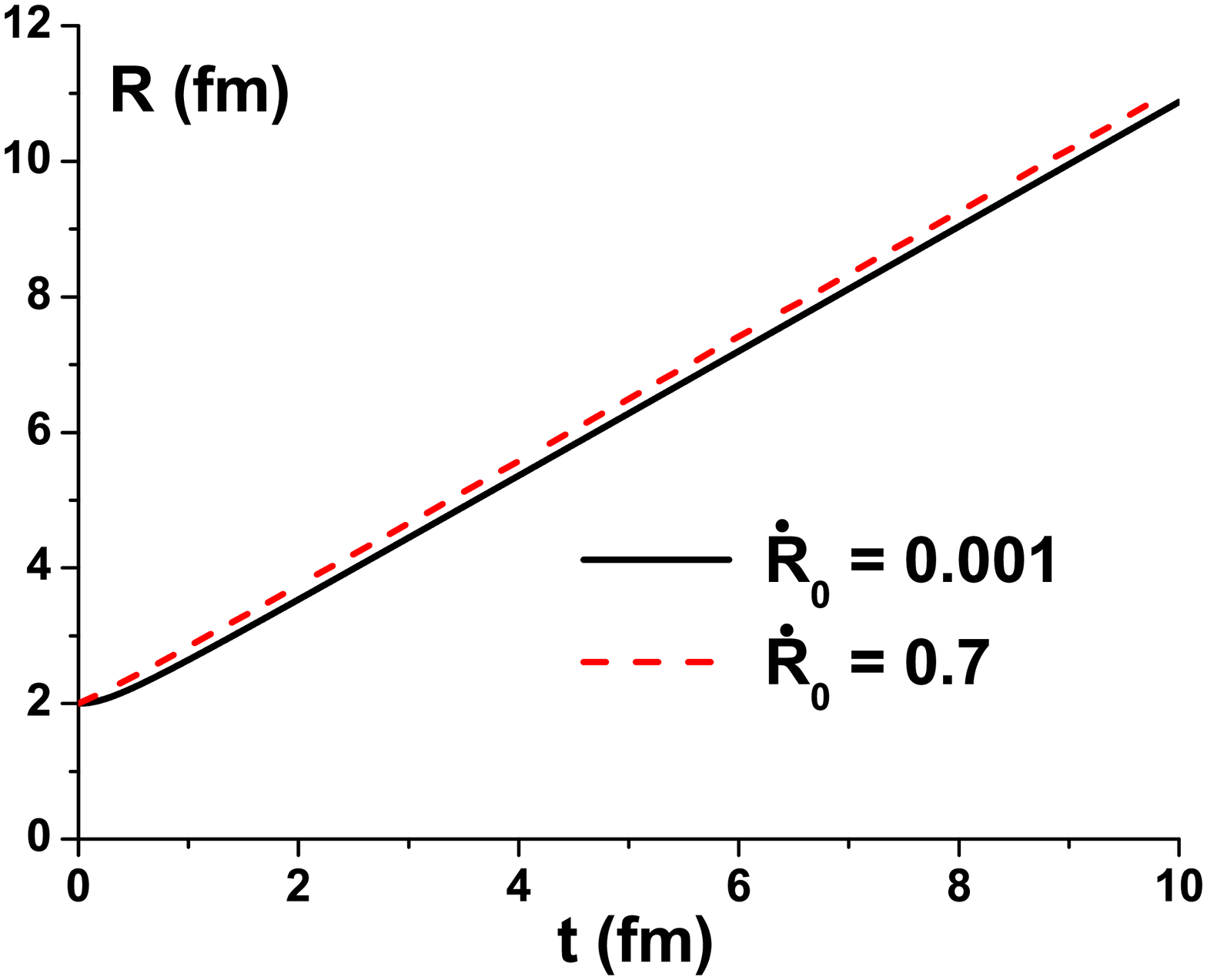}}
\subfigure[ ]{\label{fig5b}
\includegraphics[width=0.48\textwidth]{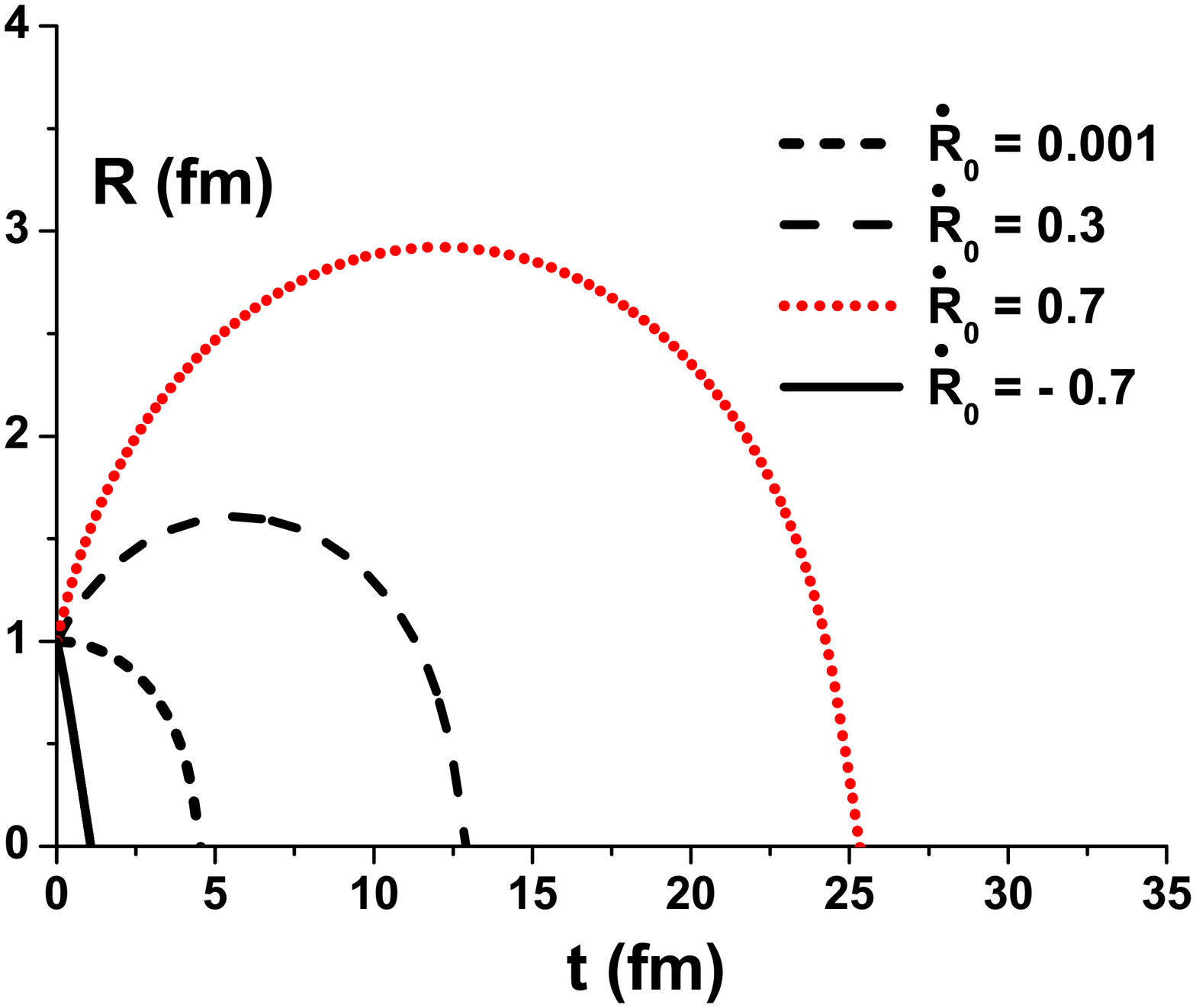}}
\subfigure[ ]{\label{fig5d}
\includegraphics[width=0.48\textwidth]{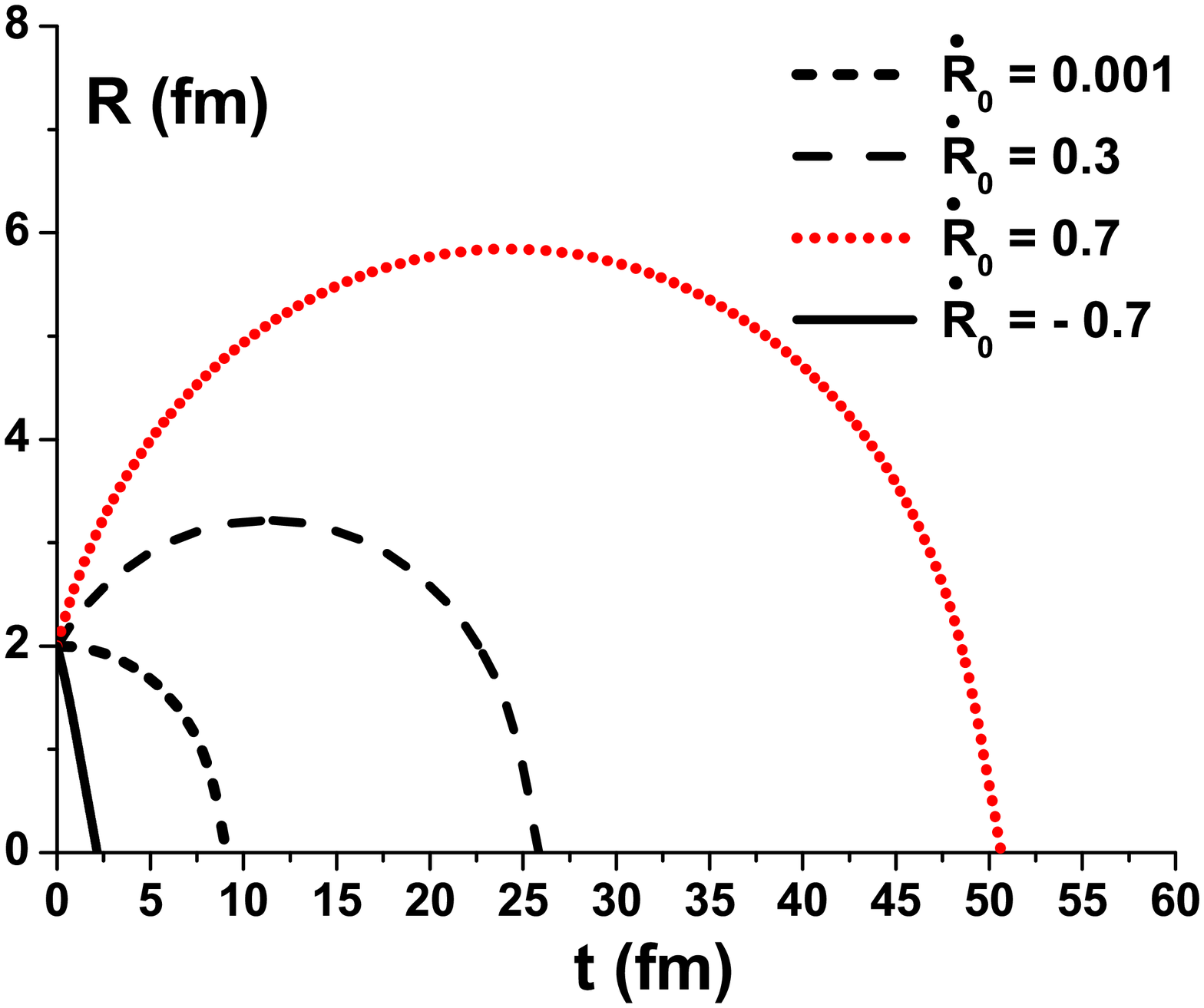}}
\end{center}
\caption{Time evolution of a bubble given by the numerical solutions of (\ref{rrpf}). QGP bubble  with an initial radius $1$ fm (a) and $2$ fm (b) expanding in a
hadron gas medium.  ($T_{ex}=140\,MeV$ and $T_{in}=120\,MeV$).  Hadron gas bubble of radius $1$ fm (c) and $2$ fm (d) surrounded by QGP  ($T_{ex}=120\,MeV$ and
$T_{in}=138\,MeV$).  }
\label{fig5}
\end{figure}

In this subsection we solve the Rayleigh-Plesset equation at finite temperature. The derivation of the new equation is in the Appendix. Essentially, the  changes
with respect to the zero temperature case can be incorporated in the  pressure, which is redefined and called $p^{eff}$.
In Fig. \ref{fig5} we present the numerical solutions of equation (\ref{rrpf}) at high chemical potential $\mu=350\,MeV$. In Figs. \ref{fig5a} and \ref{fig5c}
a QGP bubble (described by Lattice QCD) expands in a hadron gas medium (with NLWM EOS ) and in Figs. \ref{fig5b} and \ref{fig5d} a hadron gas bubble collapse
in a QGP medium. We use the two
equations of state at finite temperature presented in the previous section.  In Figs. \ref{fig5a} and \ref{fig5c} the expansion happens because
${p_{in}}^{eff}>{p_{ex}}^{eff}$  while in  Figs. \ref{fig5b} and \ref{fig5d}  the collapse occurs because is ${p_{in}}^{eff} <{p_{ex}}^{eff}$.
Comparing Figs. \ref{fig5}  and  \ref{fig3} we can conclude that, at finite temperatures, a QGP bubble in a hot hadron
gas expands faster than at zero temperature. However, even in this case, the time required for a bubble to grow and reach a radius comparable to the size of the
system ($R \simeq 6$ fm) is still large ($t \simeq 6$ fm ). On the other hand, according to (\ref{rata}) at larger temperatures the nucleation rate is larger.
At finite temperature the hadron gas to QGP transition is thus easier but the typical transition times are still large. The same considerations made for the zero
temperature case apply here. It seems very difficult to create a large region with QGP! We would really need many more bubbles. The complete conversion of the hadronic
system to a quark gluon plasma happens only after the remaining hadron bubbles collapse. In Figs. \ref{fig5b} and \ref{fig5d}
we observe how  the  HG bubble collapse occurs and how, increasing  the initial radial velocity, the bubble lives longer. In any case this collapse takes a few fm and
this gives further support to the previous conclusion that the transition takes a long time.

The fireball produced at FAIR may have a lifetime of $\tau \simeq 15 - 20 $ fm and a typical size of $R_F \simeq 10-15$ fm. Our results suggest that,
if the number of bubbles is small (one or a few) and if they have an initial radius of $1$ fm,
the system will spend a relatively long time to change from the hadronic to the QGP phase. In the most favorable scenario,
shown in Fig. \ref{fig3f} the time needed to fill the system with QGP  would be at least $10$ fm or more. Therefore in order to have a significant
fraction of the initial hadronic matter  converted to QGP,  it would be necessary to have a large number of bubbles and/or bubbles with a larger radius.
However, as shown in \cite{madsen} (for low  temperatures, $T$, and neglecting the surface tension) the formation of bubbles with a larger radius is
suppressed, with the nucleation rate  given by:
\begin{equation}
{\mathcal R} \simeq exp \{ - \frac{16 \pi}{3} \frac{\gamma \, r_c}{T} \}
\label{rata}
\end{equation}
where $\gamma$ is the curvature energy density and $r_c$ is the critical radius, i.e., the radius of the smallest bubble capable of growing, which is given by:
$$
r_c = \sqrt{\frac{2 \gamma}{(p_Q - p_H)}}
$$
where $p_Q$ and $p_H$ are the pressure in the quark and hadron phase respectively. The number of bubbles is proportional to the nucleation rate and should be larger
for bubbles with smaller radius. The inclusion of the surface tension, $\alpha$, would not help to make hadron-quark  transition faster.
Concerning this last point we note that, in a state of mechanical equilibrium between the quark bubble and the external hadronic medium,  we have the following relation
between the pressures $p_Q$ and  $p_H$  \cite{lugones}:
$$
p_Q = \frac{2 \alpha}{R} + \frac{2 \gamma}{R^2} + p_H
$$
For the quark bubble to expand, departing from equilibrium, the left side of this equality must be greater than the right side.
Nonzero values of $\alpha$ and $\gamma$ will make the expansion more difficult.

The discussion of the last paragraph is only qualitative, but is suggests that in the bubble nucleation approach to the hadron-quark phase transition, it is not so
easy to have an instantaneous phase conversion. In view of  the large bubble expansion times found in our calculations, it is possible that, even taking into account
the other aspects of nucleation,  the phase transition time is non-negligible. A large transition time implies that the matter in the fireball, even though with right
density and/or right energy density, is not immediately
in the QGP phase and hence it would not be correct to treat it with a QGP equation of state. Instead one would have to treat the system as being in  a
super dense hadronic phase or in a mixed phase.
This would be our message to realistic model makers.   Given the approximations made, this message is, for now, only a cautionary remark. In order to make a stronger
statement we would have to improve our calculations in, at least, three aspects: i) abandon the hypothesis that the external medium   is infinite and
include the effects of a  finite size fireball; ii) allow for a dependence of the pressure and energy density on the bubble radius; iii) include  surface tension and
curvature energy terms in the RRP.

\section{Conclusions}

In the early stages of relativistic heavy ion collisions at FAIR and NICA bubbles of quark gluon plasma are expected to be formed.
We have estimated the expansion time of a QGP bubble immersed in a hadron gas  solving numerically the relativistic Rayleigh-Plesset equation.
As input we have used reasonable equations of state for the  hadron and  quark-gluon plasma phases. In spite of the approximations used and the
uncertainties in the inputs, we can safely say that the expansion time is not small, being of the order of a few fm or even more. This result suggests that
the duration of the hadron-quark phase transition has to be taken into account in the simulations of the intermediate energy collisions to be performed at FAIR
and NICA. Our results depend strongly on the QGP  equation of state. The calculations were performed at zero and finite temperature, with similar conclusions.

The present work can be  extended to the study of the QGP-to-hadron gas transition in the final stage of nuclear collisions.
During the expansion it may happen that a QGP bubble survives as a supercooled domain, which eventually
disappears emitting hadrons. The dynamics of this supercooled QGP bubble was studied in detail in \cite{bub2}. Later, the effects of finite chemical
potential were discussed in \cite{bub3}.
During the fluid evolution another phenomenon may take place: the formation of a bubble of hadrons in a still hot QGP, a ``superheated'' hadron domain.
This  might happen if there is cavitation in the QGP \cite{raja,romacav,we2015a}: due to the bulk viscosity the pressure in the QGP drops
reaching a value where the existence of the hadron phase becomes possible. Once formed, the  hadron bubble suffers the pressure of the surrounding QGP  and,
depending on the pressure difference between the two phases, may expand or collapse, reforming the quark gluon plasma in that region.
All  these situations can be treated with the RRP equation.

\section{Appendix}

In natural units $\hbar=c=k_{B}=1$ the Lagrangian for a spherically symmetric system can be obtained from a variational principle and is given by \cite{rrp}:
\begin{equation}
\mathcal{L}=-4\pi \, \int_{0}^{\infty} dr \, r^{2} \, \varepsilon(n)
\label{lagra}
\end{equation}
where $\varepsilon$ is the energy density, $n$ is the density given by $n=\rho/\gamma$, where $\rho$ is the number of particles per volume (number density)
and $\gamma$ is the Lorentz factor: $\gamma=(1-v^2)^{-1/2}$ .

We consider the following scenario: a relativistic  bubble of radius $R$ at a fixed temperature immersed in a medium with other fixed temperature. These
temperatures will be introduced later in the calculations with the use of thermodynamical relations.

The Lagrangian (\ref{lagra}) is decomposed into the contributions of internal ($in$) and external ($ex$)  phases:
\begin{equation}
\mathcal{L}=\mathcal{L}_{in}+\mathcal{L}_{ex}=
-4\pi \,\int_{0}^{R} dr \, r^{2} \, \varepsilon_{in}(n_{in})
-\lim_{R_{\infty}\to\infty} \, \bigg\{ \,4\pi \,\int_{R}^{R_{\infty}} dr \, r^{2} \, \varepsilon_{ex}(n_{ex}) \, \bigg\}
\label{lagras}
\end{equation}
where the subscript $in$ refers to the matter inside the bubble and the subscript $ex$ refers to the matter outside the bubble.  For simplicity we omit the
limit of $R_{\infty}$ in the text, performing it in the end of calculations.  The number density for each phase is given by:
\begin{equation}
\rho_{in}={\frac{N_{in}}{\bigg( {\frac{4}{3}}\pi R^{3} \bigg)}}
\hspace{2cm} \textrm{and} \hspace{2cm}
\rho_{ex}={\frac{N_{ex}}{\bigg[ {\frac{4}{3}}\pi (R_{\infty}^{3}-R^{3}) \bigg]}}
\label{numberdenss}
\end{equation}
where $N_{in}$ is the number of particles inside the bubble and $N_{ex}$ is the number of particles outside the bubble.  The field velocity for each phase is
given by (\ref{fieldvg}) and (\ref{fieldvl}).  The Lorentz factors are given by (\ref{lorentzfactors}).

The relativistic version of the Rayleigh$-$Plesset equation (RRP) is obtained by the
Euler-Lagrange equation:
\begin{equation}
{\frac{d}{dt}}\bigg({\frac{\partial \mathcal{L}}{\partial\dot{R}}} \bigg)=
{\frac{\partial \mathcal{L}}{\partial R}}
\label{eulerlagrange}
\end{equation}
Inserting (\ref{lagras}) in (\ref{eulerlagrange}) yields:
\begin{equation}
{\frac{d}{dt}} \bigg[ \, \int_{0}^{R} dr \, r^{2} \,{\frac{\partial \varepsilon_{in}}{\partial \dot{R}}} +
\int_{R}^{R_{\infty}} dr \, r^{2} \,{\frac{\partial \varepsilon_{ex}}{\partial \dot{R}}} \, \bigg] =
{\frac{\partial}{\partial R}} \bigg[\, \int_{0}^{R} dr \, r^{2} \, \varepsilon_{in} + \int_{R}^{R_{\infty}} dr \, r^{2} \, \varepsilon_{ex} \, \bigg]
\label{eulerlagrangee}
\end{equation}
Recalling the Leibniz integral rule, we can calculate the two derivatives with respect to $R$ from (\ref{eulerlagrangee}):
$$
{\frac{\partial}{\partial R}}\, \int_{0}^{R} dr \, r^{2} \, \varepsilon_{in}=
\int_{0}^{R} dr \, {\frac{\partial}{\partial R}}(r^{2}\,\varepsilon_{in})+
R^{2} \, \Big(\varepsilon_{in}{\Big{|}}_{r=R}\Big) \, {\frac{\partial R}{\partial R}}
-0^{2} \, \Big(\varepsilon_{in}{\Big{|}}_{r=0}\Big) \,{\frac{\partial 0}{\partial R}}
$$
\begin{equation}
{\frac{\partial}{\partial R}}\, \int_{0}^{R} dr \, r^{2} \, \varepsilon_{in}=\int_{0}^{R} dr \, r^{2} \, {\frac{\partial \varepsilon_{in}}{\partial R}}+R^{2} \, \Big(\varepsilon_{in}{\Big{|}}_{r=R}\Big)
\label{leibniz1}
\end{equation}
and
$$
{\frac{\partial}{\partial R}}\, \int_{R}^{R_{\infty}} dr \, r^{2} \, \varepsilon_{ex}=
\int_{R}^{R_{\infty}} dr \, {\frac{\partial}{\partial R}}(r^{2}\,\varepsilon_{ex})+
{R_{\infty}}^{2} \, \Big(\varepsilon_{ex}{\Big{|}}_{r=R_{\infty}}\Big) \, {\frac{\partial R_{\infty}}{\partial R}}
-R^{2} \, \Big(\varepsilon_{ex}{\Big{|}}_{r=R}\Big) \,{\frac{\partial R}{\partial R}}
$$
\begin{equation}
{\frac{\partial}{\partial R}}\, \int_{R}^{R_{\infty}} dr \, r^{2} \,\varepsilon_{ex}=
\int_{R}^{R_{\infty}} dr \, r^{2} \, {\frac{\partial \varepsilon_{ex}}{\partial R}}-R^{2}
\, \Big(\varepsilon_{ex}{\Big{|}}_{r=R}\Big)
\label{leibniz2}
\end{equation}
Recalling that $\varepsilon_{in}=\varepsilon_{in}(n_{in})$, $\varepsilon_{ex}=\varepsilon_{ex}(n_{ex})$ and inserting (\ref{leibniz1}) and (\ref{leibniz2}) in (\ref{eulerlagrangee}) we have:
$$
{\frac{d}{dt}} \int_{0}^{R} dr \, r^{2} \,{\frac{\partial \varepsilon_{in}}{\partial n_{in}}}{\frac{\partial n_{in}}{\partial \dot{R}}} + {\frac{d}{dt}} \int_{R}^{R_{\infty}} dr \, r^{2} \,{\frac{\partial \varepsilon_{ex}}{\partial n_{ex}}}
{\frac{\partial n_{ex}}{\partial \dot{R}}}=R^{2} \, \Big(\varepsilon_{in}{\Big{|}}_{r=R}-
\varepsilon_{ex}{\Big{|}}_{r=R}\Big)
$$
\begin{equation}
+\int_{0}^{R} dr \, r^{2} \, {\frac{\partial \varepsilon_{in}}{\partial n_{in}}} {\frac{\partial n_{in}}{\partial R}}+\int_{R}^{R_{\infty}} dr \, r^{2} \, {\frac{\partial \varepsilon_{ex}}{\partial n_{ex}}}{\frac{\partial n_{ex}}{\partial R}}
\label{eulerlagrangeapplied}
\end{equation}
To proceed with the calculations at finite temperature, we use the thermodynamic relation from \cite{yagi}:
$$
d\varepsilon=Tds+\mu dn\, , \hspace{2.0cm} \textrm{where} \hspace{2.0cm} \mu={\frac{\partial \varepsilon}{\partial n}}
$$
and also the Gibbs relation $(\varepsilon+p=\mu n +Ts)$ becomes:
\begin{equation}
{\frac{\partial \varepsilon}{\partial n}}=
{\frac{(\varepsilon +p)}{n}}-{\frac{Ts}{n}}
\label{pretermorela}
\end{equation}
The entropy density is given by \cite{yagi}:
\begin{equation}
s={\frac{\partial p}{\partial T}}
\label{entropydens}
\end{equation}
By inserting (\ref{entropydens}) in (\ref{pretermorela}) we have:
\begin{equation}
{\frac{\partial \varepsilon}{\partial n}}=
{\frac{(\varepsilon +p)}{n}}-{\frac{T}{n}}\,{\frac{\partial p}{\partial T}}
\label{termorela}
\end{equation}
and this relation will introduce the temperature in our calculations.
Performing the substitution of (\ref{termorela}) in (\ref{eulerlagrangeapplied}) we have:
$$
{\frac{d}{dt}} \int_{0}^{R} dr \, r^{2} \,{\frac{(\varepsilon_{in} +p_{in})}{n_{in}}}\,{\frac{\partial n_{in}}{\partial \dot{R}}} -
{\frac{d}{dt}} \int_{0}^{R} dr \, r^{2} \,{\frac{T}{n_{in}}}\,{\frac{\partial p_{in}}{\partial T}}\,{\frac{\partial n_{in}}{\partial \dot{R}}}+
{\frac{d}{dt}} \int_{R}^{R_{\infty}} dr \, r^{2} \,{\frac{(\varepsilon_{ex} +p_{ex})}{n_{ex}}}\,
{\frac{\partial n_{ex}}{\partial \dot{R}}}
$$
$$
-
{\frac{d}{dt}} \int_{R}^{R_{\infty}} dr \, r^{2} \,{\frac{T}{n_{ex}}}\,{\frac{\partial p_{ex}}{\partial T}}\,
{\frac{\partial n_{ex}}{\partial \dot{R}}}
=R^{2} \, \Big(\varepsilon_{in}{\Big{|}}_{r=R}-
\varepsilon_{ex}{\Big{|}}_{r=R}\Big)
+\int_{0}^{R} dr \, r^{2} \,{\frac{(\varepsilon_{in} +p_{in})}{n_{in}}}\,{\frac{\partial n_{in}}{\partial R}}
$$
\begin{equation}
-\int_{0}^{R} dr \, r^{2} \,{\frac{T}{n_{in}}}\,{\frac{\partial p_{in}}{\partial T}}\,{\frac{\partial n_{in}}{\partial R}}
+\int_{R}^{R_{\infty}} dr \, r^{2} \,{\frac{(\varepsilon_{ex} +p_{ex})}{n_{ex}}}\,{\frac{\partial n_{ex}}{\partial R}}
-\int_{R}^{R_{\infty}} dr \, r^{2} \,{\frac{T}{n_{ex}}}\,{\frac{\partial p_{ex}}{\partial T}}\,{\frac{\partial n_{ex}}{\partial R}}
\label{eulerlagrangeappliedaa}
\end{equation}
From (\ref{numberdenss}) we have the following results:
\begin{equation}
{\frac{\partial n_{ex}}{\partial R}}=-{\frac{3}{R}} \, n_{ex}+\bigg(\gamma_{ex} \,{\frac{\dot{R}}{R}} \bigg)^{2} \, {\frac{{r}^{2}}{R}} \, n_{ex}
\label{ngr}
\end{equation}
\begin{equation}
{\frac{\partial n_{ex}}{\partial \dot{R}}}=-\bigg(\gamma_{ex} \,{\frac{r}{R}} \bigg)^{2} \dot{R} \, n_{ex}
\label{ngrponto}
\end{equation}
\begin{equation}
{\frac{\partial n_{in}}{\partial R}}={\frac{3R^{2}}{(R_{\infty}^{3}-R^{3})}} \, n_{in}
-{\frac{R^{3}(2R_{\infty}^{3}+R^{3})(R_{\infty}^{3}-R^{3})^{2}}
{r^{4}(R_{\infty}^{3}-R^{3})^{3}}}{(\gamma_{in} \, \dot{R})}^{2} \, n_{in}
\label{nlr}
\end{equation}
and
\begin{equation}
{\frac{\partial n_{in}}{\partial \dot{R}}}=-\bigg({\frac{\gamma_{in} \, R^{2}}{r^{2}}}\bigg)^{2}  \,
\bigg[{\frac{(R_{\infty}^{3}-r^{3})}{(R_{\infty}^{3}-R^{3})}}\bigg]^{2} \, \dot{R} \,  n_{in}
\label{nlrponto}
\end{equation}
Inserting (\ref{ngr}) to (\ref{nlrponto}) in (\ref{eulerlagrangeappliedaa}) we find that:
$$
-{\frac{d}{dt}} \, \Bigg\{ {\frac{\dot{R}}{R^{2}}}\int_{0}^{R} dr \, r^{4} \,\bigg(\varepsilon_{in} +p_{in}-T\,{\frac{\partial p_{in}}{\partial T}}\bigg){\gamma_{in}}^{2}
$$
$$
+ R^{4}\dot{R} \int_{R}^{R_{\infty}}
{\frac{dr}{r^{2}}} \,\bigg(\varepsilon_{ex} + p_{ex}-T\,{\frac{\partial p_{ex}}{\partial T}}\bigg){\gamma_{ex}}^{2}
\bigg[{\frac{(R_{\infty}^{3}-r^{3})}{(R_{\infty}^{3}-R^{3})}}\bigg]^{2}\Bigg\}
$$
$$
=R^{2} \, \Big(\varepsilon_{in}{\Big{|}}_{r=R}-
\varepsilon_{ex}{\Big{|}}_{r=R}\Big)+\int_{0}^{R} dr \, r^{2} \,\bigg(\varepsilon_{in} +p_{in}-T\,{\frac{\partial p_{in}}{\partial T}}\bigg)
\Bigg[ -{\frac{3}{R}}+\bigg(\gamma_{in} \,{\frac{\dot{R}}{R}} \bigg)^{2} \, {\frac{{r}^{2}}{R}} \Bigg]
$$
\begin{equation}
+\int_{R}^{R_{\infty}} dr \, r^{2} \,\bigg(\varepsilon_{ex} +p_{ex}-T\,{\frac{\partial p_{ex}}{\partial T}}\bigg)
\Bigg[{\frac{3R^{2}}{(R_{\infty}^{3}-R^{3})}}
-{\frac{R^{3}(2R_{\infty}^{3}+R^{3})(R_{\infty}^{3}-R^{3})^{2}}
{r^{4}(R_{\infty}^{3}-R^{3})^{3}}}{(\gamma_{ex} \, \dot{R})}^{2} \Bigg]
\label{almosteulerlagrangeapplieda}
\end{equation}
We now consider the approximation of large $R_{\infty}$ before taking the limit $R_{\infty}\rightarrow \infty$ ; this allows writing:
\begin{equation}
\bigg[{\frac{(R_{\infty}^{3}-r^{3})}{(R_{\infty}^{3}-R^{3})}}\bigg]^{2}\cong
{\frac{(R_{\infty}^{3})^{2}}{(R_{\infty}^{3})^{2}}} \cong 1
\label{aprox1}
\end{equation}
\begin{equation}
{\frac{3R^{2}}{(R_{\infty}^{3}-R^{3})}} \cong {\frac{3R^{2}}{R_{\infty}^{3}}}
\label{aprox2}
\end{equation}
and
\begin{equation}
-{\frac{R^{3}(2R_{\infty}^{3}+R^{3})(R_{\infty}^{3}-R^{3})^{2}}
{r^{4}(R_{\infty}^{3}-R^{3})^{3}}} \cong -{\frac{R^{3}(2R_{\infty}^{3})(R_{\infty}^{3})^{2}}
{r^{4}(R_{\infty}^{3})^{3}}} \cong -{\frac{2R^{3}}{r^{4}}}{\frac{(R_{\infty}^{3})^{3}}{(R_{\infty}^{3})^{3}}}  \cong -{\frac{2R^{3}}{r^{4}}}
\label{aprox3}
\end{equation}
For large $R_{\infty}$, eq. (\ref{lorentzfactors})for the Lorentz factor outside the bubble tells us that:
\begin{equation}
\gamma_{ex}=\Bigg[1-\Bigg({\frac{R^{2}(R_{\infty}^{3}-r^{3})}{r^{2}(R_{\infty}^{3}-R^{3})
\dot{R}}} \Bigg)^{2}\Bigg]^{-1/2} \cong \Bigg[1-\Bigg({\frac{R^{2}\,R_{\infty}^{3}}{r^{2}\,R_{\infty}^{3}\,\dot{R}}} \Bigg)^{2}\Bigg]^{-1/2}
=\Bigg[1-\Bigg({\frac{R^{2}}{r^{2}\,\dot{R}}} \Bigg)^{2}\Bigg]^{-1/2}
\label{lorentzfactorsaprox}
\end{equation}
We use Eqs. (\ref{aprox1}) through (\ref{lorentzfactorsaprox}) in order to rewrite Eq.  (\ref{almosteulerlagrangeapplieda}) as
$$
-{\frac{d}{dt}} \, \Bigg\{ {\frac{\dot{R}}{R^{2}}}\int_{0}^{R} dr \, r^{4} \,\bigg(\varepsilon_{in} +p_{in}-T\,{\frac{\partial p_{in}}{\partial T}}\bigg)
{\gamma_{in}}^{2}  + R^{4}\dot{R} \int_{R}^{R_{\infty}}
{\frac{dr}{r^{2}}} \,\bigg(\varepsilon_{ex} + p_{ex}-T\,{\frac{\partial p_{ex}}{\partial T}}\bigg){\gamma_{ex}}^{2}\Bigg\}
$$
$$
=R^{2} \, \Big(\varepsilon_{in}{\Big{|}}_{r=R}-
\varepsilon_{ex}{\Big{|}}_{r=R}\Big)+\int_{0}^{R} dr \, r^{2} \,\bigg(\varepsilon_{in} +p_{in}-T\,{\frac{\partial p_{in}}{\partial T}}\bigg)
\Bigg[ -{\frac{3}{R}}+\bigg(\gamma_{in} \,{\frac{\dot{R}}{R}} \bigg)^{2} \, {\frac{{r}^{2}}{R}} \Bigg]
$$
\begin{equation}
+\int_{R}^{R_{\infty}} dr \, r^{2} \,\bigg(\varepsilon_{ex} +p_{ex}-T\,{\frac{\partial p_{ex}}{\partial T}}\bigg)
\Bigg[{\frac{3R^{2}}{R_{\infty}^{3}}}
-{\frac{2R^{3}}{r^{4}}}{(\gamma_{ex} \, \dot{R})}^{2} \Bigg]
\label{almosteulerlagrangeappliedarwt}
\end{equation}
Now, following  \cite{rrp} we make the  approximation:
$$
\int_{R}^{R_{\infty}} dr \, r^{2} \,\bigg(\varepsilon_{ex} +p_{ex}-T\,{\frac{\partial p_{ex}}{\partial T}}\bigg)
{\frac{3R^{2}}{R_{\infty}^{3}}} \cong {\frac{3R^{2}}{R_{\infty}^{3}}}\,
\bigg(\varepsilon_{ex} +p_{ex}-T\,{\frac{\partial p_{ex}}{\partial T}}\bigg){\Big{|}}_{R_{\infty}}\int_{R}^{R_{\infty}} dr \, r^{2}
$$
$$
= {\frac{3R^{2}}{R_{\infty}^{3}}}\,
\bigg(\varepsilon_{ex} +p_{ex}-T\,{\frac{\partial p_{ex}}{\partial T}}\bigg){\Big{|}}_{R_{\infty}}\, \bigg({\frac{R_{\infty}^{3}}{3}}-{\frac{R^{3}}{3}} \bigg)
$$
\begin{equation}
=R^{2}\,\bigg(\varepsilon_{ex} +p_{ex}-T\,{\frac{\partial p_{ex}}{\partial T}}\bigg){\Big{|}}_{R_{\infty}}
-{\frac{R^{5}}{R_{\infty}^{3}}}\,\bigg(\varepsilon_{ex} +p_{ex}-T\,{\frac{\partial p_{ex}}{\partial T}}\bigg){\Big{|}}_{R_{\infty}}
\label{aproinfini}
\end{equation}
Inserting (\ref{aproinfini}) into (\ref{almosteulerlagrangeappliedarwt}) and
performing the limit $R_{\infty} \rightarrow \infty$  in the resulting expression, we have:
$$
-{\frac{d}{dt}} \, \Bigg\{ {\frac{\dot{R}}{R^{2}}}\int_{0}^{R} dr \, r^{4} \,\bigg(\varepsilon_{in} +p_{in}-T\,{\frac{\partial p_{in}}{\partial T}}\bigg){\gamma_{in}}^{2}  + R^{4}\dot{R} \int_{R}^{\infty}
{\frac{dr}{r^{2}}} \,\bigg(\varepsilon_{ex} + p_{ex}-T\,{\frac{\partial p_{ex}}{\partial T}}\bigg){\gamma_{ex}}^{2}\Bigg\}
$$
$$
=R^{2} \,(\varepsilon_{in}-\varepsilon_{ex}){\Big{|}}_{R}-{\frac{3}{R}}\int_{0}^{R} dr \, r^{2} \,\bigg(\varepsilon_{in} +p_{in}-T\,{\frac{\partial p_{in}}{\partial T}}\bigg)
+{\frac{{\dot{R}}^{\,2}}{R^{3}}}\int_{0}^{R} dr \, r^{4} \,\bigg(\varepsilon_{in} +p_{in}-T\,{\frac{\partial p_{in}}{\partial T}}\bigg){\gamma_{in}}^{2}
$$
\begin{equation}
+R^{2}\,\bigg(\varepsilon_{ex} +p_{ex}-T\,{\frac{\partial p_{ex}}{\partial T}}\bigg){\Big{|}}_{\infty}
-2R^{3}{\dot{R}}^{\,2}\int_{R}^{\infty} \, {\frac{dr}{r^{2}}} \,\bigg(\varepsilon_{ex} +p_{ex}-T\,{\frac{\partial p_{ex}}{\partial T}}\bigg)
{\gamma_{ex}}^{2}
\label{eulerlagrangelimited}
\end{equation}
Introducing the new variable $x=r/R$ it is possible to rewrite the integrals which contain Lorentz factors and then (\ref{eulerlagrangelimited}) becomes:
$$
-{\frac{d}{dt}} \, \Bigg\{ R^{3}\dot{R} \Bigg[ \int_{0}^{1} dx \, x^{4} \,\bigg(\varepsilon_{in} +p_{in}-T\,
{\frac{\partial p_{in}}{\partial T}}\bigg){\gamma_{in}}^{2}  + \int_{1}^{\infty}
{\frac{dx}{x^{2}}} \,\bigg(\varepsilon_{ex} + p_{ex}-T\,{\frac{\partial p_{ex}}{\partial T}}\bigg){\gamma_{ex}}^{2} \Bigg] \Bigg\}=
$$
$$
= R^{2} \, \Bigg[ (\varepsilon_{in}-\varepsilon_{ex}){\Big{|}}_{R}+\bigg(\varepsilon_{ex} +p_{ex}-T\,{\frac{\partial p_{ex}}{\partial T}}\bigg){\Big{|}}_{\infty} \,
\Bigg]-{\frac{3}{R}}\int_{0}^{R} dr \, r^{2} \,\bigg(\varepsilon_{in} +p_{in}-T\,{\frac{\partial p_{in}}{\partial T}}\bigg)
$$
\begin{equation}
+R^{2}\, {\dot{R}}^{\,2} \Bigg[ \int_{0}^{1} dx \, x^{4} \,\bigg(\varepsilon_{in} +p_{in}-T\,{\frac{\partial p_{in}}{\partial T}}\bigg){\gamma_{in}}^{2}
-2 \int_{1}^{\infty}{\frac{dx}{x^{2}}} \,\bigg(\varepsilon_{ex} + p_{ex}-T\,{\frac{\partial p_{ex}}{\partial T}}\bigg){\gamma_{ex}}^{2} \Bigg]
\label{almeulerlagrangelimited}
\end{equation}
in which we can identify the integrals (\ref{prei1}) and (\ref{prei2}), the effective pressures (\ref{effectives}) and also the function $F$ given by (\ref{efep}).
With these already determined quantities, Eq. (\ref{almeulerlagrangelimited}) is the
Relativistic Rayleigh-Plesset equation that appears in (1).

\begin{acknowledgments}
This work was partially financed by the Brazilian funding agencies CAPES, CNPq and FAPESP. We thank Takeshi Kodama and Jorge Horvath for instructive discussions.
\end{acknowledgments}

\end{document}